\documentclass[twocolumn,amsmath,amssymb,aps,pra,superscriptaddress,showpacs,10pt]{revtex4}

\usepackage{graphicx}% Include figure files

\usepackage{bbm}% bold math
\usepackage{color}
\usepackage[utf8]{inputenc}
\usepackage[T1]{fontenc}

\newcommand{\bra}[1]{\left\langle #1 \hspace{0.2ex} \right|} %bra
\newcommand{\ket}[1]{\left| \hspace{0.2ex} #1 \right\rangle} %ket 

\newcommand{\half}{\frac{1}{2}}

\begin{document}
\title{Spectral properties of finite laser-driven lattices of ultracold Rydberg atoms}

% \date{\today}

\author{Nikolas Tezak}
\affiliation{Zentrum f\"ur Optische Quantentechnologien, Universit\"at Hamburg, Luruper Chaussee 149, 22761 Hamburg, Germany}
\affiliation{Physikalisches Institut, Universit\"at Heidelberg, Philosophenweg 12, 69120 Heidelberg, Germany}

\author{Michael Mayle}
\altaffiliation[Present address: ]{JILA, University of Colorado and National Institut of Standards and Technology, Boulder, Colorado 80309-0440, USA}
\affiliation{Zentrum f\"ur Optische Quantentechnologien, Universit\"at Hamburg, Luruper Chaussee 149, 22761 Hamburg, Germany}

\author{Peter Schmelcher}
\email[]{peter.schmelcher@physnet.uni-hamburg.de}
\affiliation{Zentrum f\"ur Optische Quantentechnologien, Universit\"at Hamburg, Luruper Chaussee 149, 22761 Hamburg, Germany}

\date{\today}% It is always \today, today,

\begin{abstract}\label{txt:abstract}
We investigate the spectral properties of a finite laser-driven lattice of ultracold Rydberg atoms exploiting the dipole blockade effect
in the frozen Rydberg gas regime.  Uniform one-dimensional lattices as well as lattices with variable spacings are considered.
In the case of a weak laser coupling, we find a multitude of many-body Rydberg states with well-defined excitation properties which are adiabatically
accessible starting from the ground state. A comprehensive analysis of the degeneracies of the spectrum as well as of the single and pair excitations numbers of
the eigenstates is performed. In the strong laser regime, analytical solutions for the pseudo-fermionic eigenmodes are derived. 
Perturbative energy corrections for this approximative approach are provided. 
\end{abstract}

\maketitle
%%%%%%%%%%%%%%%%\left[ \left\lbrace %%%%%%%%%%%%%%%%%%%%%\section{Introduction}%%%%%%%%%%%%%%%%%%%%%%%%%%%%%%%%%%%%%%%%%%%%%%%%%%%%%%%%%%%%%%%%%%%%%%%%%%%%%%%%%%%%%%%%%
\section{Introduction} % (fold)
\label{sec:introduction}
In the past decades, experimental investigations of ultracold atoms have reached an unprecedented control over
the motional degree of freedom as well as the interaction of the atoms. One of the intriguing systems in this field are ultracold Rydberg atoms 
since they exhibit strong long-range interactions of isotropic or even anisotropic character.
In fact, the interaction-induced level shift of collective states featuring multiple Rydberg excitations can be large enough to exceed the excitation laser linewidth thereby preventing the excitation of further Rydberg atoms. 
This dipole blockade effect was predicted by theory one decade ago along with a proposal how it could be exploited for the realization of fast two-qubit quantum gates \cite{PhysRevLett.85.2208,PhysRevLett.87.037901}. Since then, a wide range of more general quantum information processing applications has emerged, putting Rydberg atoms in the focus of many theoretical as well as experimental efforts (see
\cite{RevModPhys.82.2313} and references therein). Experimentalists have found the dipole blockade effect 
\cite{tong:063001,singer:163001,liebisch:253002,vogt:073002,ditzhuijzen:243201} in gases of alkali atoms and recently for
just two individually trapped atoms \cite{Urban2009,Gaetan2009}. The coherence of the observed effects
was demonstrated \cite{Deiglmayr2006293, heidemann:163601,  raitzsch:2008, cooperyounge:2009} and the resulting
collective Rabi-frequency \cite{heidemann:163601} has been observed.

The difficulties often encountered in working with a large number of atoms are related to the spatial structure of their gaseous samples.
Because of the varying particle density typically encountered in a trap and the spatially varying intensity profile of the excitation laser, no unique collective Rabi-frequency is encountered. Several theory groups worked to circumvent these limitations by performing simulations of the excitation dynamics in unordered samples looking for signatures of the coherent dynamics \cite{hernandez:2008a, hernandez:2008b,PhysRevA.76.013413} and predicting the formation of crystalline structures \cite{Pohl:2010p1627}.
In \cite{PhysRevLett.98.023002} it was proposed to use a laser detuned from the single atom resonance to actively produce
Rydberg atom pairs with the specific distance that corresponds to an interaction induced energy shift equal to the detuning. This
effect was observed indirectly by measuring the interaction-induced ionization rates as a function of the interaction
time \cite{PhysRevLett.104.013001}. In spite of these successes, it remains clear that a spatially ordered Rydberg gas is highly
desirable. To this end, both theorists as well as experimentalists have concentrated on identifying a stable trapping mechanism for Rydberg states that allows a strong confinement in magnetic \cite{choi:243001,lesanovsky:053001,hezel:223001,mayle:113004,mayle:053410} as well as optical \cite{PhysRevLett.104.173001,1367-2630-12-2-023031,dutta00} traps. While the creation of lattice traps is straightforward nowadays in the optical regime, ongoing experimental effort is put into creating arrays of magnetic traps \cite{gerritsma:033408,Whitlock2009}.

First theoretical investigations of the Rydberg excitation in structured ultracold atomic gases \cite{Olmos:2008p1776} consider a ring-shaped lattice of ground state atoms, concentrating on the case where the laser coupling to the Rydberg state is weak in comparison to the next-neighbour Rydberg interaction. In the opposite regime of a dominant laser coupling, the same authors demonstrated that the
system permits the formation of fermionic collective excitations \cite{Olmos:2009p285}.

In the present work, we consider a finite one-dimensional lattice of ground state atoms that are coherently excited to the Rydberg state via a two-photon laser transition. We provide a thorough investigation of the spectral properties of this system for single-spaced as well as variably-spaced lattices. Different parameter regimes are covered.
The model Hamiltonian that forms the basis of our investigations is derived in Sec.~\ref{sec:model_hamiltonian}.
In Section \ref{sec:weak_laser_regime}, the weak laser regime is analyzed where we find that a multitude of system states with well-defined excitation properties are adiabatically accessible from the ground state. Among these are the crystalline states that have been previously discussed in \cite{Pohl:2010p1627}. In Section \ref{sec:strong_laser_regime}, we focus on the strong laser regime where we achieve an approximate description through the XY-model of the antiferromagnetic spin chain \cite{Lieb1961407}. We provide analytical solutions for the pseudo-fermionic eigenmodes as well as the perturbative energy corrections that are necessary due to the approximate nature of the description. The intermediate regime between a weak and strong laser coupling is briefly treated numerically in Sec.~\ref{sec:intermediate_regime}.
In Sec.~\ref{sec:alternating_spacings} we consider the case of multiple lattice spacings and describe in Sec.~\ref{sec:perfect_crystal_transitions} a specific spatial setup that allows the emulation of the crystalline state transitions of a larger sized lattice. 
The appendix contains a derivation of the perturbative energy corrections in the strong laser regime.

\section{The Model} % (fold)
\label{sec:model_hamiltonian}

Starting from a quite general model including all individual atoms contributions, one can derive a simplified effective Hamiltonian that describes the system in terms of a lattice of two-level systems. To this end, the internal structure of an individual atom is described in terms of a ground state $g$, an excited (Rydberg) state $e$, and an intermediate state $m$ that we include in order to allow for an experimentally realistic excitation scheme via a two-laser setup: $g \longleftrightarrow m \longleftrightarrow e$. After transforming into a rotating frame of reference and a subsequent coarse-graining in time, an effective, static Hamiltonian is obtained and the intermediate $m$ state can be removed from the state space by an adiabatic elimination if the excitation lasers are strongly detuned with respect to this level \cite{scully97,0953-4075-43-15-155003}.  After this, the two-laser setup is modelled by an effective $g \longleftrightarrow e$ coupling and the description at each lattice site $k$ can be further reduced by employing the superatom states $\ket{e}_{k}=(N_k)^{-1/2}\sum_i^{N_k}\ket{g^{(k)}_1,g^{(k)}_2,\dots, e^{(k)}_i,\dots,g^{(k)}_{N_k}}$ and $\ket{g}_{k}=\ket{g^{(k)}_1,g^{(k)}_2,\dots,g^{(k)}_{N_k}}$, respectively.
Here $N_k$ is the number of atoms on site $k$. This means that either all atoms of a given site are in the ground state or they symmetrically share a single excitation \cite{heidemann:163601}. This reduction is justified only if the confinement to a single lattice site is sufficiently strong such that all atoms are located within a distance of each other that is smaller than the dipole blockade radius and also much smaller
than the lattice constant. The symmetrization of the excited superatom state presupposes that both the effective coupling strength of the two-photon transition and its detuning from resonance are identical for all atoms within a given site. By considering solely the internal level structure of each atom, the frozen Rydberg gas regime is presumed where the atomic centre of mass motion can be neglected on the timescale of the coherent excitation dynamics.

The above considerations lead to our model Hamiltonian
\begin{align}\label{eq:hamiltonian}
H &=  \sum_{k=1}^N \left[ \half \Omega^{(k)} \sigma_x^{(k)} + \half \Delta^{(k)}
  \sigma_z^{(k)} + \sum_{l=k+1}^{N} \mathcal{V}_{kl}
  n_e^{(k)}n_e^{(l)}\right].
\end{align}
Here, the operators $\sigma^{(k)}_i, i \in \{x,y,z\}$ act on the superatom located at site $k$ and take on the
usual Pauli-matrix form when expressed in the local superatom basis $\mathcal{S}^ {(k)} := \left\{\ket{e}_{k},\ket{g}_{k}\right\}$. The excitation number operators may also be expressed in terms of the Pauli-operators, $n_e^{(k)} = \frac{1}{2} [ \sigma_z^{(k)} + \mathbbm{1}] = \ket{e}_k\bra{e}_k$. The system is therefore formally equivalent to a spin-1/2 lattice with interactions $\mathcal{V}_{kl}$. In this picture,
the contributions due to the laser are similar to the interaction of the spins with an external magnetic field with a (local) component $\Delta^{(k)}$ aligned with the spins and a perpendicular component $\Omega^{(k)}$.

Hamiltonian (\ref{eq:hamiltonian}) contains three different contributions.
First, the laser coupling of each single atom's ground state to the excited state is given by
\begin{align}\label{eq:HL}
	H_L = \half \sum_{k=1}^N   \Omega^{(k)} \sigma_x^{(k)},
\end{align}
where $\Omega^{(k)}:= \sqrt{N_k}\Omega_0$ denotes the collective Rabi frequency for the superatom state at site $k$ \cite{heidemann:163601}.
Because of the number of atoms $N_k$ contributing to the excitation dynamics, $\Omega^{(k)}$ is enhanced by a factor of $\sqrt{N_k}$ compared to the single atom Rabi frequency $\Omega_0$.
Second, the part which describes the site-dependent laser detuning,
\begin{align}\label{eq:HD}
	H_D = \half \sum_{k=1}^N  \Delta^{(k)} \sigma_z^{(k)},
\end{align}
corresponds to an energy gap (in the effective RWA picture) of $\Delta^{(k)}$ between the excited state and the ground state of the superatom at site $k$.
Finally, the Rydberg interactions read as pairwise interactions between each two sites $k,l$ with an interaction strength that depends on the spatial separation of the sites,
\begin{align}\label{eq:Hint}
H_{\text{int}} = \sum_{k=1}^N \sum_{l=k+1}^{N} \mathcal{V}_{kl} n_e^{(k)}n_e^{(l)}.
\end{align}
Each summand is non-zero only if both affected sites are in the
excited state. We only consider repulsive interactions, i.e.,
$\mathcal{V}_{kl}\ge 0$, which are common for interacting Rydberg atoms
in their $ns$-state for a wide range of principal quantum numbers $n$ \cite{0953-4075-38-2-021}.

The contributions (\ref{eq:HL}-\ref{eq:Hint}) may be classified by two separate
criteria: By locality, i.e., whether or not they act non-trivially on
more than a single site, and by their simultaneous diagonalizability.
The laser contributions, i.e., the laser coupling $H_L$ and the laser detuning $H_D$ are local, while the Rydberg interactions $H_{\text{int}}$ are by definition non-local.
On the other hand, the Rydberg interactions and the laser detuning operator commute, $[H_D,H_{\text{int}}] = 0$,
since they can both be expressed in terms of $\sigma_z^{(k)}$-operators and the identity.
We thus identify two interesting limiting parameter regimes:
In the \emph{weak laser coupling regime} the Hamiltonian is dominated by its diagonal contributions: the Rydberg interactions and the laser detuning.
Alternatively, in the \emph{strong laser regime} the Hamiltonian is dominated by local operators. As demonstrated
in \cite{Olmos:2009p285} for a ring lattice, this allows for an approximation of the system by an XY-model.
Unless stated differently, we will restrict ourselves to
global laser parameters $(\Omega^{(k)},\Delta^{(k)})\longrightarrow
(\Omega,\Delta)$. 

We describe the system in terms of the 'canonical' product
basis 
\begin{align}\label{eq:canonical_product_basis}
\mathcal{S}_N:=\left\{ \ket{s_1s_2\dots s_N},\quad s_k \in \{e,g\}
\right\},
\end{align}
since these states are directly accessible in experiments. 
Moreover, our Hamiltonian is already diagonal in this basis except
for the laser coupling part $H_L$.
Calculating the matrix elements of our Hamiltonian $H$ for the canonical product
basis $\mathcal{S}_N$ yields a sparse matrix. Specifically, it is
straightforward to show that the number of non-zero matrix elements is
given by $(N+1)2^N = D\log_2 2D$, where $D = \#\mathcal{S}_N = 2^N$ is the dimension of
the state space. 
To simplify our notation, we define the canonical product ground state and the fully excited state as 
$\ket{G} := \ket{gg \dots g}$ and $\ket{E} := \ket{ee \dots e}$, respectively.

\section{Single-spaced Lattices} % (fold)
\label{sec:single_spaced_lattices}

We start by considering lattices with a single lattice spacing $a$.
In this case, the interaction potential is given by
\begin{align}
	\mathcal{V}_{kl} = V_{|l-k|} := \frac{C_n}{a^n |l-k|^n} = \frac{V_1}{|l-k|^n}.
\end{align}
We will usually assume a Van-der-Waals interaction potential, i.e., $n=6$.
The restriction to global laser parameters yields the final Hamiltonian
\begin{align}\label{eq:master_hamiltonian_constant_couplings}
	H & = \half \Omega \sum_{k} \sigma_x^{(k)} + \half \Delta \sum_{k} \sigma_z^{(k)} + V_1 \sum_{l>k} \frac{n_e^{(k)}n_e^{(l)}}{|l-k|^n}.
\end{align}
This Hamiltonian is invariant under reflections at the centre of the lattice. As was done for the ring lattice in \cite{Olmos:2009p285}
we designate the corresponding unitary operator for
this symmetry as $\mathcal{R}$ and define it via its action on the
local Pauli operators,
\begin{align*}
  \mathcal{R}^{\dagger}\sigma_{n}^{(k)}\mathcal{R} = \sigma_{n}^{(N-k+1)},
\end{align*}
for $n = x,y,z$. The product ground state is invariant under reflections $\mathcal{R}\ket{G} = \ket{G}$ and since the full basis can be constructed by means of the ground state and the Pauli-operators, this completely determines the form of $\mathcal{R}$.
Clearly, reflecting the system twice should leave it unchanged and hence $\mathcal{R}^{\dagger} = \mathcal{R}^{-1} = \mathcal{R}$.
The eigenvalues of $\mathcal{R}$ are thus given by $\pm 1$.

For a weak laser coupling $|\Omega| \ll V_1, |\Delta|$, the diagonal
contributions to the Hamiltonian dominate. In this case the laser
coupling leads to a small off-diagonal perturbation. 
Alternatively, in the case of weak Rydberg interactions 
the Hamiltonian can be mapped approximately to an XY-model Hamiltonian.
In the following, both limiting regimes are discussed in detail and analytical formulas describing the excitation spectra are derived. 
For the intermediate regime, where neither of the above conditions is fulfilled, numerical simulations reveal the full spectrum.

\subsection{Weak Laser Regime}
\label{sec:weak_laser_regime}
We start by considering the weak laser regime, i.e., assuming $|\Omega| \ll V_1, |\Delta| $.
The Hamiltonian (\ref{eq:master_hamiltonian_constant_couplings}) is conveniently divided into two parts, $H  = H_0 + H'$,
grouping together the laser detuning with the next-neighbour Rydberg interactions to give the dominant contribution
\begin{align}\label{eq:H0}
	H_0  &  = \half \Delta \sum_k \sigma_z^{(k)} + V_1 \sum_{k=1}^{N-1} n_e^{(k)}n_e^{(k+1)},
\end{align}
while the perturbation consists of the laser coupling as well as the long range Rydberg interactions,
\begin{align}
	H'  &  = \half \Omega \sum_k \sigma_x^{(k)} + \frac{V_1}{2^n} \sum_{d = 2}^{N-1} \frac{1}{(d/2)^n } \sum_{k = 1}^{N-d} n_e^{(k)} n_e^{(k+d)},
\end{align}
which we have rewritten as a sum of contributions for a given separation $d$.
Introducing the operators for the total excitation number,
\begin{align}
    N_e & = \sum_{k=1}^N n_e^{(k)} =  \frac{1}{2} \sum_{k=1}^N \sigma_z^{(k)} + \frac{N}{2},
\end{align}
and the next-neighbour excitation pair number, 
\begin{align}
N_{ee} = \sum_{k=1}^{N-1} n_e^{(k)}n_e^{(k+1)},
\end{align}
we can rewrite the unperturbed Hamiltonian as $	H_0  =  \Delta \left(  N_e - N/2 \right)+ V_1 N_{ee}$.
For a given state $\ket{S} = \ket{s_1s_2\dots s_N} \in
\mathcal{S}_N$ from the canonical product basis [cf.~\eqref{eq:canonical_product_basis}], the unperturbed
energy eigenvalue is thus given by
\begin{align}\label{eq:energy_ne_nee}
	E(S) = \Delta [N_e(S) - N/2] + V_1 N_{ee}(S),
\end{align}
where $N_e(S)$ and $N_{ee}(S)$ denote the eigenvalues of the
operators $N_e$ and $N_{ee}$ for the state $\ket{S}$. 
They can readily be obtained by counting the number of
excitations $\dots e \dots$ and excitation bonds $\dots ee \dots$ present in
the sequence $S = s_1s_2\dots s_N$. There are several important observations to be made at this point: 
\begin{enumerate}
	\item $H_0$ is linear in $N_e$ and $N_{ee}$. Since $N_e$ and
          $N_{ee}$ are diagonal, this linear energy relation holds for
          their eigenvalues as well.
	\item In general, there are multiple canonical product states of equal $(N_e, N_{ee})$ which are consequently always degenerate with respect to the unperturbed Hamiltonian. 
	\item Depending on the specific ratio of $\Delta$ and $V_1$, the simple form of \eqref{eq:energy_ne_nee} already suggests that additional degeneracies between states of different $(N_e, N_{ee})$ are possible. Since degeneracies in the RWA picture correspond to resonant laser couplings in the non-rotating frame, our result simply states that we can tune the laser to resonantly excite multi-particle states, as one would intuitively expect.
\end{enumerate}

We define $D_N(N_e,N_{ee})$ to be the dimension of each 
$(N_e, N_{ee})$-subspace $\mathcal{H}_{(N_{e},N_{ee})}$. The calculation of $D_N(N_e,N_{ee})$ is
possible through a combinatorial analysis. 
First, note that for a given state $\ket{S} = \ket{s_1s_2\dots s_N}$ the number of next-neighbour
excitation pairs $N_{ee}(S)$ is fully determined by the total number of excitations $N_e(S)$ and the
number of excited domains $ee\dots e$ within $\ket{S}$ which we denote by
$d_e(S)$. Whenever $d_e(S)$ is equal to one, $N_{ee}$ takes
on its maximal value $\left. N_{ee} \right|_{d_e=1} = N_e - 1$. For
each additional domain $N_{ee}$ decreases by one if $N_e$
remains fixed. Hence, the following relation holds for any $N_e, N_{ee}$:
\begin{align}
   d_e(S) = N_e(S) - N_{ee}(S).
\end{align}
We can now calculate $D_N$ by analyzing the number of possibilities of how to construct appropriate sequences '$s_1s_2\dots s_N$'.
For $N_e = 0$ we can only have $\ket{S} = \ket{gg \dots g} = \ket{G}$, hence we assume $N_e, d_e \ge 1$.
We must distribute $N_e$ excitations across $d_e \le N_e $ domains 
which leads to a factor of $\binom{N_e - 1}{d_e -1} = \binom{N_e - 1}{N_e-N_{ee} -1}$. Now, for any such division of the excited states into domains, we must count the number of ways how to distribute these domains of excited atoms across the lattice such that there is always at least one ground level site between two excited domains. The number of ground level sites is given by $N_g = N - N_e$. The domains of excited sites can thus be inserted at $N_g -1 $ positions between two '$g$'-characters or at the two positions at either end of the lattice. Hence, there are $N_g + 1$ positions across which we distribute $d_e$ excited domains and we must multiply the above result by a factor of $\binom{N_g+1}{d_e} = \binom{N - N_e + 1}{N_e - N_{ee}}$.
Together, we find
\begin{align}
    D_N(N_e\ge1, N_{ee}) = \binom{N_e - 1}{N_e-N_{ee} -1}\binom{N - N_e + 1}{N_e - N_{ee}}
\end{align}
and $D_N(0,N_{ee}) = \delta_{N_{ee},0}$.

 \begin{figure}
 	\centering
 	\includegraphics[width=8cm]{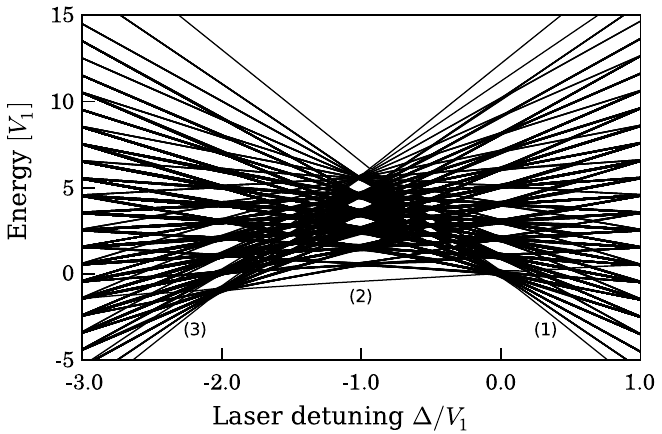}
 	\caption{\label{fig:spectra_weak_omega_varying_delta} Energy
          spectrum for $N=13$ lattice sites and a constant laser
          coupling $\Omega = 0.05 V_1$. At specific rational values
          of $\Delta/V_1$ many energy levels are almost
          degenerate. Depending on $\Delta/V_1$ the energetic ground
          state in the RWA frame is given by $(1)$ the canonical
          product ground state, $(2)$ the alternating state
          $\ket{ege\dots ge}$, and $(3)$ the fully excited state.}
 \end{figure}
The spectrum of Hamiltonian (\ref{eq:H0}), i.e., the full set of unperturbed energy eigenvalues as a function of a varying ratio $\Delta/V_1$ exhibits points of high degeneracy, see figure \ref{fig:spectra_weak_omega_varying_delta}. These occur only at specific rational values of
$\Delta/V_1$. In principle, these can be calculated from
\eqref{eq:energy_ne_nee} if one takes into consideration the
combinatorially possible combinations of $N_e$ and $N_{ee}$. For
$\Delta=0$ the lasers are tuned to atomic resonance. In this case the
state $\ket{gg\dots g}$ is degenerate (within the effective
  RWA picture) with all states that lack neighbouring
excitations which corresponds to dipole-blocleqed states. However, for
non-zero detuning the state $\ket{gg\dots g}$ can also be brought to
degeneracy with other states. Physically, this corresponds to a
situation in which the lasers resonantly couple $\ket{gg\dots g}$ to
states containing pairs of neighbouring excitations. Note, however, that due to
the local nature of the laser coupling Hamiltonian $H_L$,
these couplings require the presence of intermediate, in general off-resonant
states. A state from a given $(N_e, N_{ee})$ subspace can be coupled via $H_L$ to states from $(N_e \pm 1, N_{ee})$, $(N_e \pm 1, N_{ee} \pm 1)$ and $(N_e \pm 1, N_{ee} \pm 2)$. 

A particular example is provided by $\Delta/V_1 = -1/2$: In this case $\ket{gg\dots g}$ is degenerate with states containing exactly two neighbouring excitations $\left\{ \ket{eeg\dots g},\ket{geeg\dots g},\dots \ket{g\dots gee}\right\}$.
The general condition for a degeneracy between two canonical product states $\ket{S_1}$ and $\ket{S_2}$ is given by
\begin{align} \label{eq:degeneracy_condition}
	 \frac{\Delta}{V_1} & = -\frac{N_{ee}(S_1)  - N_{ee}(S_2)}{N_e(S_1) - N_e(S_2)}.
\end{align}
For $N_{e}(S_1) \neq N_{e}(S_2)$ and $N_{ee}(S_1) \neq N_{ee}(S_2)$ the equation can only be solved if the ratio $\Delta/V_1$ is rational because the eigenvalues of $N_e$ and $N_{ee}$ are integral.
A combinatorial analysis reveals that the quantum numbers $N_e$ and $N_{ee}$ obey the following constraints in addition to being integral:
\begin{align} \label{eq:ne_nee_constraints_1}
	N_e = 0 & \Rightarrow N_{ee} = 0, \\ \label{eq:ne_nee_constraints_2}
	1 \le N_e \le \lceil N /2 \rceil & \Rightarrow 0 \le N_{ee} \le N_e - 1 , \\ \label{eq:ne_nee_constraints_3}
	\lceil N /2 \rceil < N_e & \Rightarrow 2N_e - N - 1 \le N_{ee} \le N_e-1.
\end{align}
These render it difficult to provide a concise, general quantitative analysis of the possible degeneracies.

\begin{figure*}
      \centering
      \includegraphics[width=16cm]{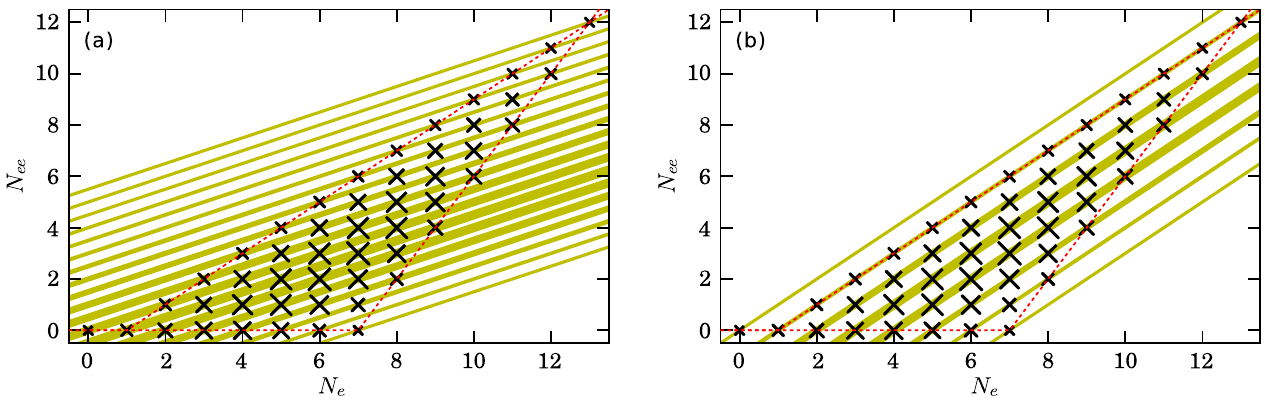}
      \caption{ \label{fig:ne_nee_pairs} Lines of constant energy
        (yellow) in the $(N_e,N_{ee})$ plane for $N=13$ and
        for different slopes (a) $\Delta/V_1=-1/2$ and (b) $\Delta/V_1=-1$, respectively. The line thickness indicates the dimension of the
        associated energy eigenspace. The dashed red lines illustrate the
        constraints [see (\ref{eq:ne_nee_constraints_1},\ref{eq:ne_nee_constraints_2},\ref{eq:ne_nee_constraints_3})].
        The size of an $X$-marker indicates the
        dimension $D_N(N_e, N_{ee})$ of the corresponding subspace. The minimal number
      of non-degenerate energy levels is attained for $\Delta
      = -V_1$, cf.\ subfigure (b). Note that in this
      case the canonical product ground state $(N_e=0, N_{ee}=0)$ is
      energetically isolated.}
 \end{figure*}
\begin{figure}
      \centering
\includegraphics[width=8cm]{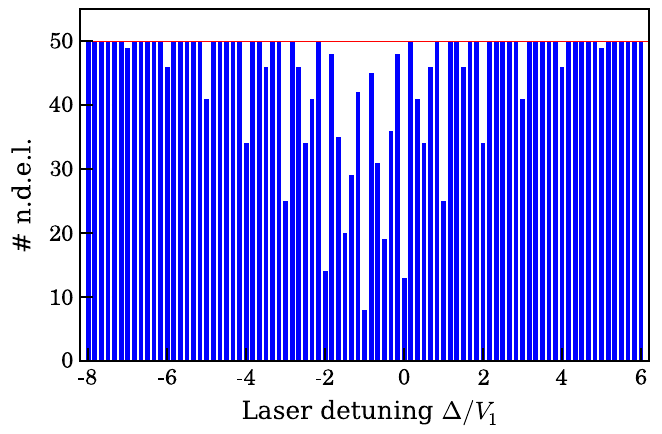}
      \caption{\label{fig:degeneracy_slope}The number $\xi$ of
        non-degenerate energy levels for different ratios $\frac{\Delta}{V_1}$ for $N=13$. For any $N$, the minimum number is attained for $\Delta = - V_1$. In our particular case we find $\xi_\text{min}=8$ and $\xi_\text{max}=50$, respectively.}
\end{figure}

However, to gain a better understanding of the spectrum and its degeneracies we can visualize the combinatorial possibilities for $(N_e, N_{ee})$ and try to recognize general principles that are independent of the lattice size. To this end, figure \ref{fig:ne_nee_pairs} illustrates all possible pairs $(N_e, N_{ee})$ for $N= 13$.
For a fixed ratio $\Delta/V_1$, (\ref{eq:energy_ne_nee}) can
also be interpreted as describing the lines of constant energy in the
$(N_e, N_{ee})$ plane. These lines can be
constructed by drawing a line of slope $\lambda = -{ \Delta /
  V_1}$ through each valid $(N_e, N_{ee})$ point. A degeneracy occurs
whenever two or more points lie on the same line.
The number $\xi$ of remaining non-degenerate energy levels depends sensitively on the ratio $\Delta /V_1$.
The minimal such number $\xi_\text{min}$, i.e., the highest degeneracy is attained for $\Delta /V_1 = -1$.
This is true independently of $N$ since it follows from the shape of
the distribution of possible $(N_e, N_{ee})$ pairs. 
However, the actual value of $\xi_\text{min}$ depends on the system size and is given by $\xi=\lceil N/2\rceil+1$.
The maximal value of $\xi$, on the other hand, is determined by the possible $(N_e,N_{ee})$ pairs for a fixed number of lattice sites and reads $\xi_\text{max}=1 + \lceil N/2\rceil^2$.
Results of a systematic calculation of the number of non-degenerate energy levels for many different ratios ${\Delta/ V_1}$ are shown in Figure~\ref{fig:degeneracy_slope} for $N=13$.
When moving away from the point of the highest level of degeneracy ($\Delta /V_1 = -1$) the number of non-degenerate energy levels increases from its minimal value $\xi_\text{min}=8$ and rapidly reaches its maximal value $\xi_\text{max} = 50$, except for the
points of integer $\Delta/V_1$ for which pronounced local minima of $\xi$ are encountered.

In the following, we proceed by analyzing the most relevant parameter regimes in more detail. We are especially interested in the level structure of the energy band that includes the canonical product ground state $\ket{G}$ or the fully excited state $\ket{E}$. For a strongly blue detuned laser $\Delta/V_1 \approx -2$, for example, we find that the fully excited state $\ket{E}$ is degenerate with all other states that fulfil $N_{ee} = 2N_{e} - N - 1 \ge 0$; the latter are given by the right hand side of the red 'triangle' of states compatible with the constraints, cf.~figure \ref{fig:ne_nee_pairs}. These states minimize the next-neighbour Rydberg interaction energy for a given number of excitations $N_{e}$.
Another interesting parameter setting is given by $\Delta /V_1 \approx -1/2$. This leads to the resonant creation of isolated pairs of neighbouring excitations. Specifically, the subspace that includes the product ground state $\ket{G}$ is characterized by the relation $2N_{ee} = N_{e}$.

\subsubsection{The Full Blockade Regime}
\label{sec:the_full_blockade_regime}
The full blockade regime is attained by restricting both the laser detuning and the laser coupling to be very small compared to the next-neighbour Rydberg interaction energy. The relevant subspace that includes the canonical product ground state is then approximately (neglecting the mixing due to a non-zero $\Omega$) given by the 0-eigenspace of $N_{ee}$, i.e., all states that contain no neighbouring excitations as given by the lowest row in figure \ref{fig:ne_nee_pairs}. Note that this implies $N_e \le \lceil N/2 \rceil$.

For non-zero detuning, the full-blockade subspace is energetically separated from the remaining state space if the detuning fulfils
$|\Delta| < V_1 / N$. Introducing the generalized excitation pair number operators
\begin{align}
	N_{ee}^{[l]} = \sum_{k = 1}^{N-l} n_e^{(k)} n_e^{(k+l)},
\end{align}
we may rewrite our full Hamiltonian as
\begin{align}\label{eq:hamiltonian_grouped_interactions_by_range}
	H  = \frac{\Delta}{2} \sum_{k= 1}^N \sigma_z^{(k)}  + \frac{\Omega}{2} \sum_{k= 1}^N \sigma_x^{(k)} + V_1 \sum_{l = 1}^{N-1} \frac{1}{l^n}N_{ee}^{ [l] }.
\end{align}
For a positive detuning $\Delta>0$, the canonical product ground state is clearly also the ground state of the diagonal part of Hamiltonian (\ref{eq:hamiltonian_grouped_interactions_by_range}). However, at $\Delta = 0$, even when considering the part of the Rydberg interactions beyond nearest neighbours, this state becomes degenerate with all $(N_e = 1, N_{ee} = 0)$ states. For small negative, i.e., blue detuning, these states form a degenerate ground state energy level, up until $\Delta/V_1 = 1/(N-1)^n$ at which point they cross with the $N_e = 2$ state $\ket{eg\dots ge}$. Further decreasing the detuning leads to a succession of ground states with an increasing number of excitations $2 \le N_e\le \lceil N/2 \rceil$. In general the ground state energy level is degenerate, except when $(N-1)/(N_e-1) = l_{\rm min} \ge 2$ is integral:
In this case, it is possible to evenly distribute the excitations across the lattice. hence, there exists a unique minimal energy \emph{crystal state} $\ket{[N_e]} := \ket{eg\dots geg\dots ge}$ where the spacing between neighbouring excitations is given by $l_{\rm min}$. This state's eigenvalue for the generalized excitation pair operators is given by
\begin{align}
N_{ee}^{[l]} \ket{[N_e]} = \begin{cases} (N_e-q)\ket{[N_e]} & \text{ for } q = l/l_{\rm min} \in \mathbb{N}, \\ 0  & \text{ otherwise.} \end{cases}
\end{align}
The corresponding diagonal matrix elements of the Hamiltonian accordingly read
\begin{align}
    \bra{[N_e]}H\ket{[N_e]}_d&  = \Delta (N_e-N/2) + \frac{V_1}{l_{\rm
        min}^n} \sum_{q=1}^{N_e - 1}\frac{N_e - q}{q^n}\\
& \approx \Delta (N_e-N/2) + \frac{V_1}{l_{\rm
        min}^n} (N_e-1) \\
& =  \Delta (N_e-N/2) +  \frac{(N_e-1)^{n+1}}{(N-1)^n} V_1.
\end{align}
When $(N-1)/(N_e-1)$ is non-integral, the minimal energy
configurations are realized by states with excitations separated by at
least two different spacings. This case is more complex to analyze in
full generality.

If both $N_e$, $N_e-1\in
\mathbb{N}_{\ge2}$ are divisors of $N-1$, the crystal states
$\ket{[N_e+1]}, \ket{[N_e]}$ feature an avoided
level-crossing for a well-defined detuning $\Delta_{N_e}^{N_e +1}$:
\begin{align} \label{eq:transition_delta_exact}
 - \frac{\Delta_{N_e}^{N_e +1}}{V_1} &  =
 \left(\frac{N_e}{N-1}\right)^n  \left\{
   \sum_{k=1}^{N_e}\frac{ 1}{k^n} \right.  \\ \nonumber
    & \quad  + \left. \left[1 -\left( 1- \frac{1}{N_e}\right)^n \right]
 \sum_{k=1}^{N_e-1} \frac{N_e - k}{k^n} +  \frac{1}{N_e^n} \right\}. \\
&  \approx  \frac{N_e^{n+1}- (N_e-1)^{n+1}}{(N-1)^n}.
\end{align}
In the thermodynamic limit $N \to \infty$, $N_e/N \to f \in (0,1/2]$
of the exact expression \eqref{eq:transition_delta_exact} the
results of Weimer et al.\ are reproduced,
$- \tilde{\Delta}/V_1 \to f^n \left[ (n+1) \zeta(n) + O(1/N_e)\right]$ \cite{weimer:2010}.
For our modest lattice sizes this limit is not properly realized, but
we can see that the scaling of the transition detunings with the
excitation fraction $f = N_e/N$ is determined by the scaling of the
interaction potential with the distance. For van der Waals interactions (i.e.,
$n=6$) this implies that the transition detunings for small excitation
fractions are very small.

\begin{figure}
 	\centering
        \includegraphics[width=8cm]{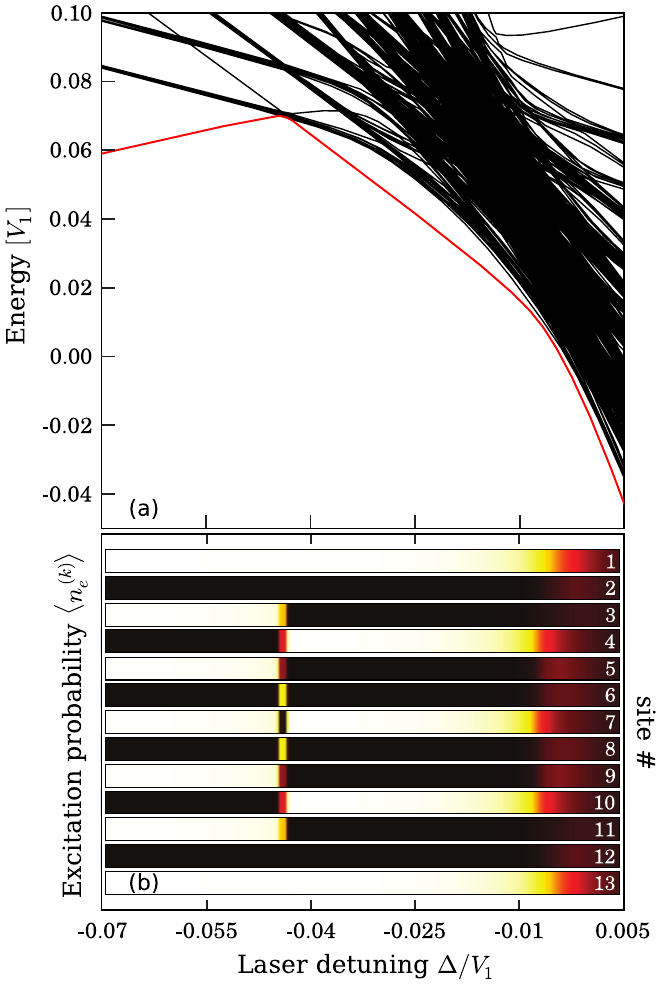}
 	\caption{\label{fig:spectra_state_analysis_full_blockade}
(a) Detail of the energy spectrum near the ground state for $N=13$ lattice sites in the full blockade regime for small detunings and very small laser coupling $\Omega/V_1 = 0.005$. The red line indicates the ground state. At specific detunings, given by (\ref{eq:transition_delta_exact}), avoided crossings are encountered where the ground state changes from one crystal state to another. 
(b) Site-resolved local excitation probability of the ground state as a function of the same laser detuning as in subfigure (a). We see from right to left a sequence of states featuring an increasing number of evenly spaced excitations. White regions correspond to unit Rydberg excitation probability, whereas dark regions denote the ground state.}
 \end{figure}

Whenever one is interested in the crystal states $\ket{[N_e]}$,
it is favourable to choose a lattice size $N$ for which many different
$\ket{[N_e]}\to \ket{[N_e+1]}$ transitions can be realized. 
For example, one can verify that $N=13$ allows for
the transitions $\ket{[2]}\to\ket{[3]}\to\ket{[4]}\to\ket{[5]}$. The next candidate of
an ideal lattice size is given by $N=61$ which allows for the additional transitions $\ket{[5]}\to\ket{[6]}\to\ket{[7]}$.
In Section \ref{sec:patterned_lattices} we present an alternative approach how to realize these transitions by means of a non-uniform lattice with a specific pattern of lattice spacings.
Figure~\ref{fig:spectra_state_analysis_full_blockade} illustrates the
succession of crystal states for the above-mentioned case of $N=13$.
An interesting aspect about this phenomenon is that although the
locations of the level crossings, i.e., the specific values of the
detuning at which they occur, are determined by the scaling of the
interaction potential with the distance, the existence of such
crystal ground states already follows from any kind of long-ranged repulsive
interactions.

\subsubsection{Fully Excited Lattices} \label{sec:fill_lattice}
We conclude our discussion of the weak laser regime by considering a strongly blue detuned laser $\Delta/V_1 \approx -2$. In this case, we find that the fully excited state $\ket{E}$ is degenerate with all other states that fulfil $N_{ee} = 2N_{e} - N - 1 \ge 0$. In Figure \ref{fig:ne_nee_pairs} these states make up the right hand edge of the triangle of states compatible with the constraints. 

\begin{figure}
 	\centering
        \includegraphics[width=8cm]{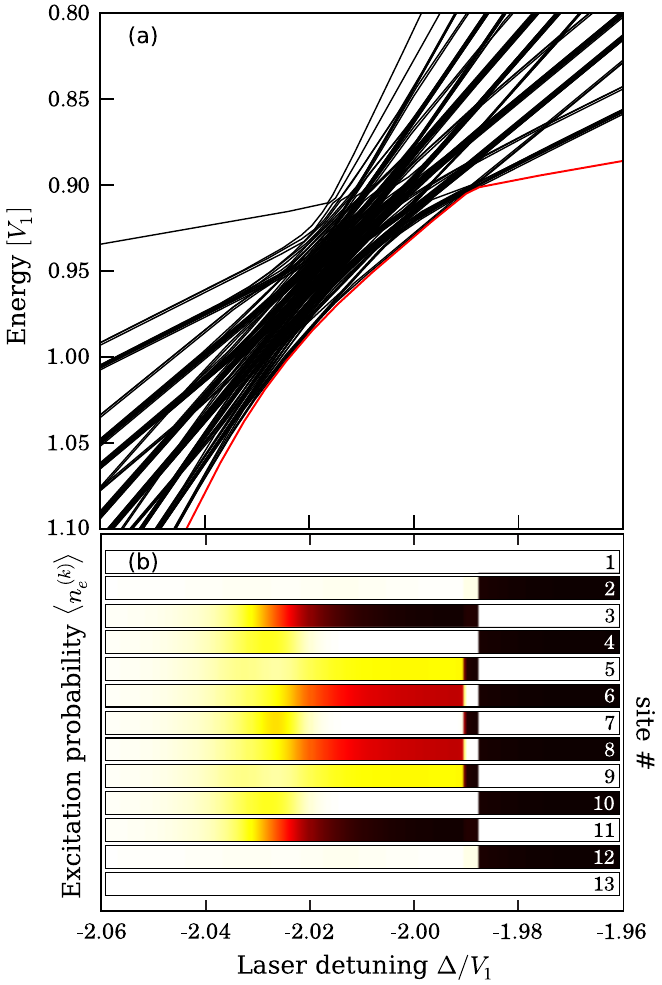}
 	\caption{\label{fig:spectra_state_analysis_filling_up} 
Same as in figure \ref{fig:spectra_state_analysis_full_blockade} but for detunings close to $\Delta/V_1=-2$. One encounters the 
          transition from the alternating excited state $\ket{ege\dots ge}$ (which is the final state is figure \ref{fig:spectra_state_analysis_full_blockade}) to the fully excited state $\ket{E}$.}
 \end{figure}

As we have seen in figure \ref{fig:spectra_weak_omega_varying_delta} for a
strongly blue detuned laser, $\Delta \le - 2 V_1$, the fully excited
state $\ket{ee \dots e}$ actually becomes energetically favoured in the
RWA frame. 
Because of the long-ranged Rydberg-Rydberg interactions one finds a succession of states with an increasing number of Rydberg excitations for increasing negative detuning. This trend continues until the completely excited lattice is encountered. Hence, the $\Delta/V_1 \approx -2$ transition concludes what was started at the $\Delta = 0$ degeneracy: a transition from the canonical product ground state to the maximally excited state 
with all sites occupied by Rydberg excitations.
The $\Delta/V_1 \approx -2$ transition is illustrated in figure \ref{fig:spectra_state_analysis_filling_up}.

\subsubsection{Resonant Pair Creation} \label{sec:anti_blockade}
\begin{figure}
 	\centering
	\includegraphics[width=8cm]{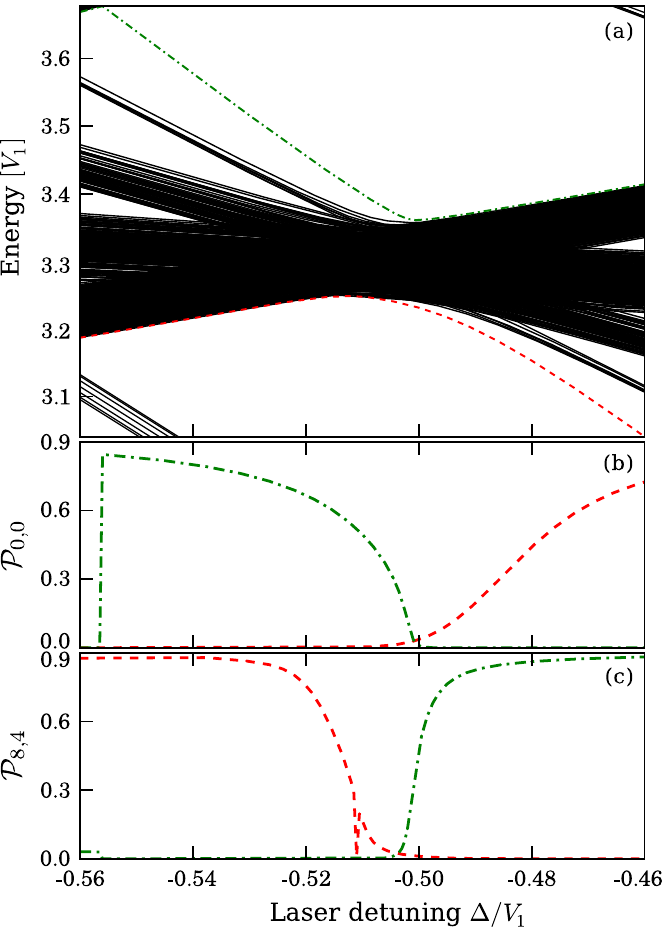}
 	\caption{\label{fig:spectra_state_analysis_anti_blockade}
(a) Detail of the energy spectrum including the canonical ground state $\ket{G}$ for detunings close to $\Delta/V_1=-1/2$, $N=13$,  and for $\Omega/V_1=0.1$.
This specific detuning favours the excitation of Rydberg pairs. In subfigures (b) and (c) the projections $\mathcal{P}_{0,0}$ and $\mathcal{P}_{8,4}$ are shown for the two distinct states marked by the dash-dotted green and dashed red line.}
 \end{figure}

As mentioned before, another interesting degeneracy is realized for
$\Delta/V_1 = -1/2$. From the degeneracy condition
\eqref{eq:degeneracy_condition} one finds that in this case the 
state $\ket{G}$ $(N_e = N_{ee} = 0)$ is degenerate with all other canonical
product states whose quantum numbers satisfy $N_e = 2 N_{ee}$.
Figure \ref{fig:spectra_state_analysis_anti_blockade}(a) presents the
spectrum at the point of degeneracy for $N=13$. The
maximal number of excitations compatible with both the combinatorial
constraints and  $N_e = 2 N_{ee}$ is given by the maximal integer
satisfying $N_{e} \le \frac{3}{2}(N-1)$. For $N=13$ this is $N_{e} =
8$ and consequently $N_{ee} = 4$.
Correspondingly, in subfigures \ref{fig:spectra_state_analysis_anti_blockade}(b) and (c) the projections $\mathcal{P}_{0,0}$ and $\mathcal{P}_{8,4}$ are shown for the two distinct states that evolve from and into the canonical ground state $\ket{G}$, respectively. $\mathcal{P}_{N_{e},N_{ee}}$ is the projector onto the subspace $\mathcal{H}_{N_{e},N_{ee}}$ containing $N_e$ Rydberg excitations and $N_{ee}$ Rydberg pairs. As expected, one finds a transition of the adiabatic eigenstates from the canonical ground state to the pair excited states when going through the region of degeneracy. At the crossing itself, further subspaces contribute to the investigated states as $\mathcal{P}_{0,0}+\mathcal{P}_{8,4}\neq 1$. The main such contribution stems from the $\mathcal{H}_{6,3}$ subspace while the $\mathcal{H}_{4,2}$ and $\mathcal{H}_{2,1}$ subspaces only contribute marginally.

\subsection{Strong Laser Regime}
\label{sec:strong_laser_regime}
As for the weak laser regime, when the laser energy contributions are very large in comparison to
the next-neighbour Rydberg interactions, the system again becomes
analytically treatable. Neglecting all long-range interactions, we rewrite the interaction part in this regime as
\begin{align}
  H_{\rm int}& \approx \frac{V_1}{4}\sum_{k =  1}^{N-1} \sigma_z^{(k)}
  \sigma_z^{(k+1)}  + \frac{V_1}{2}\sum_{k = 1}^{N} \sigma_z^{(k)}  \nonumber\\
& \qquad - \frac{V_1}{4}\left[\sigma_z^{(1)}+ \sigma_z^{(N)}\right] 
 +(N-1)\frac{V_1}{4}.
\end{align}
The total Hamiltonian can then be separated according to 
$H = H_0 + H' + H_b$
into a local part
\begin{align}
  H_0 = \half \sum_{k=1}^N\left[\Omega\sigma_x^{(k)} +
    \left(\Delta + V_1\right)\sigma_z^{(k)}\right] +(N-1)\frac{V_1}{4},
\end{align}
an interaction part
$H'  = \frac{V_1}{4}\sum_{k =  1}^{N-1} \sigma_z^{(k)} \sigma_z^{(k+1)}$,
and a boundary term $H_b = - \frac{V_1}{4}\left(\sigma_z^{(1)}+  \sigma_z^{(N)}\right)$. 
Defining the pseudo laser coupling $\tilde{\Omega} := \sqrt{\Omega^2 + (\Delta + V_1)^2} \gg V_1$,  
the local part can be re-expressed as $ H_0  := \half \tilde{\Omega} \sum_{k=1}^N \vec{n}\vec{\sigma}^{(k)} \label{eq:laser_dominant_hamiltonian}$
where we have introduced the notation $  \vec{n}\vec{\sigma}^{(k)} =
\sum_{i \in \{x,y,z\}} n_i \sigma_i^{(k)}$.
The $n_i$ are given by
\begin{align}
\vec n := \frac{1}{\tilde{\Omega}}\begin{pmatrix} \Omega \\ 0 \\
  \Delta + V_1 \end{pmatrix} = \begin{pmatrix} \sin\theta \\ 0 \\
  \cos\theta \end{pmatrix} \Rightarrow {\vec n}^2 = 1.
\end{align}
The pseudo angle $\theta \in [0, 2\pi]$ satisfies the relation
\begin{align} \label{eq:pseudo_angle_coupling_relation}
	\cos\theta = \frac{\Delta  + V_1}{\tilde{\Omega}}, \; \sin\theta = \frac{\Omega}{\tilde{\Omega}}.
\end{align}
Note that $H_0$ contains no information about the spatial arrangement of the lattice sites. It is
symmetric under any permutation of the lattice sites.
Therefore, it can be diagonalized by rotating each site $k$ independently in spin-space about the $\sigma_y^{(k)}$ axis with the unitary transformation
\begin{align}
	& U_{\theta'} := \bigotimes_{k=1}^N  \exp\left[-\frac{i}{2} \theta' \sigma_y^{(k)}\right].
\end{align}
It follows that 
\begin{align}
	U^\dagger_{\theta'} \vec{n}\vec{\sigma}^{(k)} U_{\theta'} &=
\cos(\theta-\theta') \sigma_z^{(k)} +
        \sin(\theta-\theta')\sigma_x^{(k)}.
\end{align}
Choosing $\theta'=\theta$, the transformed local Hamiltonian is given by
\begin{align} \label{eq:rotated_laser_parts}
  \tilde{H}_0 = U^\dagger_\theta H_0 U_\theta =
  \half\tilde{\Omega}\sum_{k=1}^N \sigma_z^{(k)} +(N-1)\frac{V_1}{4}.
\end{align}

\begin{figure}
      \centering
 \includegraphics[width=8cm]{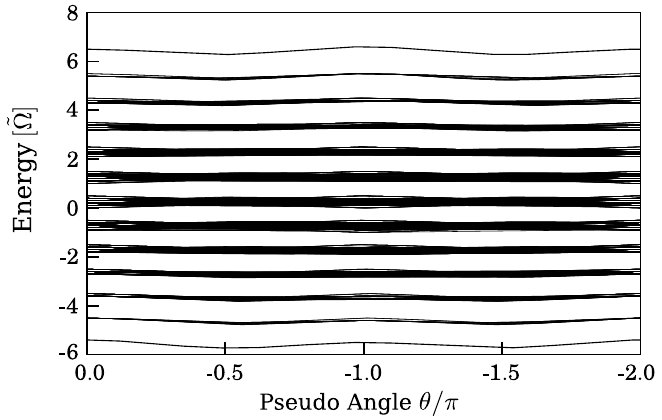}
      \caption{ \label{fig:laser_dominant_varying_theta_2}
                Energies as a function of the pseudoangle $\theta$ for $V_1 = 0.1 \tilde{\Omega}$. 
		The deviations from the constant energy level behaviour are strongest for $\Omega = 0$
		(corresponding to $\theta = 0,\pi, 2\pi$). They are always positive, which is due
		to the fact that the perturbing Rydberg interactions operator is
		strictly positive.
        }
 \end{figure}

From Hamiltonian (\ref{eq:rotated_laser_parts}), the spectral features can be straightforwardly deduced. Its eigenstates are given by products of the rotated single site basis states
\begin{align}
  \ket{+;\theta}_k & =  \cos\frac{\theta}{2} \ket{e}_k +
  \sin\frac{\theta}{2} \ket{g}_k, \\
  \ket{-;\theta}_k & =  -\sin\frac{\theta}{2} \ket{e}_k +
  \cos\frac{\theta}{2} \ket{g}_k.
\end{align}
The transformed product ground state and the fully excited state read $\ket{\tilde{G};\theta} = U_{\theta}\ket{G} = \ket{-- \dots -;\theta}$, and $\ket{\tilde{E};\theta} = U_{\theta}\ket{E} = \ket{++ \dots +;\theta}$, respectively.
Due to the full permutation symmetry of $H_0$ the
associated energy depends only on the number $\tilde{N}_+$ of $\ket{+;\theta}$
factors present in a given product state,
\begin{align}
\tilde{N}_+ & := \frac{1}{2}\sum_{k=1}^N \left[\sigma_z^{(k)} +
  1\right] = U_\theta^\dagger N_+ U_\theta,
\end{align}
where $N_+ = \frac{1}{2}\sum_{k=1}^N [\cos\theta \sigma_z^{(k)} +
  \sin\theta \sigma_x^{(k)} + 1]$.
The dominant part $H_0$ of the Hamiltonian in the strong laser coupling regime can be expressed in terms of $N_+$ and $\tilde{N}_+$ as 
\begin{equation}
 H_0 =\tilde{\Omega}(N_+ - N/2)+(N-1)\frac{V_1}{4}
\end{equation}
and
\begin{equation}
 \tilde H_0 =\tilde{\Omega}(\tilde{N}_+ - N/2)+(N-1)\frac{V_1}{4},
\end{equation}
respectively.
The eigenvalues of $N_+$ are $k=0,1,\dots, N$ and for each eigenvalue $0 \le k \le N$ the
eigenspace has dimension $\binom{N}{k}$ due to the number of
possibilities to distribute $k$ excitations across a lattice of length
$N$.
Hence, the eigenvalues of $H_0$ depend only on $\tilde{\Omega}$. Moreover, its
spectrum is symmetrical with respect to $E=0$ for $V_1=0$ and is given by 
\begin{align}
 E_\kappa=-(N-2\kappa)\tilde{\Omega}/2,\, \kappa=0,1,\dots,N.
\end{align}
This fact is illustrated in figure \ref{fig:laser_dominant_varying_theta_2},
where we confirm the constant spacing between the energy bands. The slight variation of the
energy levels with $\theta$ stems from the Rydberg interactions which are incorporated in $H'$ and $H_b$.

To understand the substructure of the energy levels at other values
for $\theta$, we must focus our attention on the interaction part $H'$.
Applying the unitary transform for $\theta' = \theta$ to the interaction Hamiltonian yields
\begin{align}
\tilde{H}' & =  \frac{V_1}{4} \cos^2\theta\sum_{k=1}^{N-1}
\sigma_z^{(k)}\sigma_z^{(k+1)} + \frac{V_1}{4} \sin^2\theta\sum_{k=1}^{N-1}
\sigma_x^{(k)}\sigma_x^{(k+1)} \nonumber \\
&\quad -\frac{V_1}{4}\sin \theta \cos \theta \sum_{k=1}^{N-1} \left[
  \sigma_z^{(k)}\sigma_x^{(k+1)} +
  \sigma_x^{(k)}\sigma_z^{(k+1)}\right]. 
\end{align}
As pointed out previously, $\theta' = \theta \approx 0, \pi$ corresponds to an
almost vanishing laser intensity, i.e., we recover the weak laser
regime (cf.~Sec.~\ref{sec:weak_laser_regime}). Consequently, all
off-diagonal terms (i.e., those featuring at least one
$\sigma_x^{(k)}$ operator as a factor) vanish due to the factor of
$\sin \theta$. Hence, within the vicinity of these values, the energy
level structure can be fully understood in terms of our previous discussion of the weak laser regime.
The conceptually opposite case is realized for $\theta' = \theta \approx \pi/2, 3\pi/2$.
In this case the laser coupling is maximal, its magnitude being given by the pseudo laser coupling which we assume to be much larger than the Rydberg interactions and the laser detuning, $|\Omega| = \tilde{\Omega} \gg V_1 = - \Delta$.
In the following we fix $\theta = \pi/2$; the case $\theta = 3\pi/2$ can be treated equivalently.
As mentioned before, this closely corresponds to the system analyzed 
in \cite{PhysRevA.81.023604}. The main differences are given by our requirement that $\Delta = - V_1$ and our open boundary conditions.
Therefore we will present in the following discussion only the major differences.

For $\theta = \pi/2$ our interaction Hamiltonian $\tilde{H}' = \frac{V_1}{4} \sum_{k=1}^{N-1}
\sigma_x^{(k)}\sigma_x^{(k+1)}$ assumes a purely off-diagonal form. The transformed boundary term accordingly reads $\tilde{H}_b = \frac{V_1}{4}[\sigma_x^{(1)} + \sigma_x^{(N)}]$.
Clearly, $\tilde{H}'$ is not diagonal in the rotated
basis, but it is possible to further separate it into a part that
commutes with $\tilde{H}_0$ and one that does not. This is achieved by
rewriting 
$\sigma_x^{(k)} =  \sigma_+^{(k)} + \sigma_-^{(k)}$, leading to
\begin{align} \label{eq:split_rotated_perturbation}
  \tilde{H}' ={} & \frac{V_1}{4} \sum_{k=1}^{N-1}
\left[\sigma_+^{(k)}\sigma_-^{(k+1)}+
  \sigma_-^{(k)}\sigma_+^{(k+1)} \right] \nonumber \\
& + \frac{V_1}{4} \sum_{k=1}^{N-1} \left[\sigma_+^{(k)}\sigma_+^{(k+1)}+
 \sigma_-^{(k)}\sigma_-^{(k+1)}\right]\nonumber\\
\equiv{} & \tilde{H}'_1+\tilde{H}'_2.
\end{align}
$\tilde{H}'_1$ consists of
operators which conserve the total number of excitations
$\tilde{N}_+$ and therefore commutes with the unperturbed Hamiltonian, i.e.,
$\left[\tilde{H}_0 ,  \sigma_+^{(k)}\sigma_-^{(k+1)} \right] =
\left[\tilde{H}_0 ,  \sigma_-^{(k)}\sigma_+^{(k+1)} \right] = 0$ for any $k\in \{1,\dots N-1\}$ . 
This implies that we can find a simultaneous eigenbasis for $\tilde{H}_0$ and $\tilde{H}'_1$.
As we will show later, the remaining part of the perturbation is fully off-diagonal in this basis. 
Hence, for our parameter regime $\tilde{\Omega}\gg V_1,\Delta$ the
dynamics of any initial state that belongs to a certain $\tilde{N}_+$
eigenspace is influenced by the off-diagonal part $\tilde{H}'_2$ only
at second order $O(V_1^2/\tilde{\Omega})$. Formally, this can be demonstrated
by the application of quasi-degenerate
perturbation theory \cite{shavitt:5711}.
We thus neglect $\tilde{H}'_2$ in the following and proceed by diagonalizing the resulting
Hamiltonian
\begin{align}
  \tilde{H}_{xy} & := \tilde{H}_0 + \frac{V_1}{4} \sum_{k=1}^{N-1}
\left[\sigma_+^{(k)}\sigma_-^{(k+1)}+
  \sigma_-^{(k)}\sigma_+^{(k+1)} \right] \\ \label{eq:xy_hamiltonian}
 & = \tilde{\Omega}\sum_{k=1}^N \sigma_+^{(k)}\sigma_-^{(k)} 
 - N\tilde{\Omega}/2  +(N-1)\frac{V_1}{4}\\
  &\qquad +\frac{V_1}{4} \sum_{k=1}^{N-1}
\left[\sigma_+^{(k)}\sigma_-^{(k+1)}+
  \sigma_-^{(k)}\sigma_+^{(k+1)} \right] \nonumber.
\end{align}
As already pointed out in \cite{Olmos:2008p1776}, (\ref{eq:xy_hamiltonian}) is the familiar
Hamiltonian of the XY-Model \cite{Lieb1961407}, which can readily be diagonalized by the introduction of fermionic ladder operators
\begin{align}
c_k :=
\sigma_-^{(k)}e^{i\pi\sum_{j=1}^{k-1}\sigma_+^{(j)}\sigma_-^{(j)}}
 =(-1)^{k-1}\sigma_-^{(k)}\prod_{j=1}^{k-1}\sigma_z^{(j)}.
\end{align}
These non-local operators and their adjoint operators obey fermionic anti-commutation rules
\begin{align}
  \{c_k,c_l\} = \{c^\dagger_k,c^\dagger_l\}  = 0,\quad  \{c_k,c^\dagger_l\}  = \delta_{kl}.
\end{align}
By repeated action of different $c_k^\dagger$ on the ground state
$\ket{\tilde{G}}$ we can construct a basis for the
state space which, by construction, is also an eigenbasis of $\tilde{N}_+$ (and therefore $\tilde{H}_0$). 
It is also straightforward to confirm the following relations:
\begin{align}
\sigma_+^{(k)}\sigma_-^{(k)} &= c_k^\dagger c_k, \\
\sigma_+^{(k)}\sigma_-^{(k+1)} &= c_k^\dagger c_{k+1}, \\
\sigma_-^{(k)}\sigma_+^{(k+1)} &= -c_k c_{k+1}^\dagger =
 c_{k+1}^\dagger c_k.
\end{align}
Hence, the XY-Hamiltonian, which is a quadratic form in the $\sigma_\pm$ operators,
becomes a quadratic form in the fermionic operators,
\begin{align}
  \tilde{H}_{xy} &= 2\tilde{\Omega} \sum_{k=1}^N c_k^\dagger c_k +
  \frac{V_1}{4}\sum_{k=1}^{N-1} \left( c_k^\dagger c_{k+1} +
    c_{k+1}^\dagger c_k\right)\\ 
&\quad -N\tilde{\Omega}/2+(N-1)\frac{V_1}{4},
\end{align}
which, by introducing the real, symmetric matrix $M = (M_{jk})_{j,k=1}^N$, we may rewrite as 
\begin{align}
  \tilde{H}_{xy} =  \sum_{j=1}^N\sum_{k=1}^N c_j^\dagger M_{jk} c_k -N\tilde{\Omega}/2.
\end{align}
The elements of $M$ are given by $M_{jk} =
\tilde{\Omega}\delta_{jk}  + \frac{V_1}{4}(\delta_{j,k+1} +
\delta_{j+1,k})$. 
By diagonalizing $M=R^T\Lambda R$, $\Lambda_{jk} =
\lambda_j\delta_{jk}$ with an orthogonal matrix $R$,
the Hamiltonian is further simplified. This corresponds to a principal
component analysis of the quadratic form in the fermionic operators.

Defining new fermionic operators $\eta_j := \sum_{k=1}^N R_{jk}c_k$,
the Hamiltonian decouples into $N$ independent 'fermionic' modes:
\begin{align}
  \tilde{H}_{\rm xy} = \sum_{k= 1}^N \lambda_k \eta_k^\dagger \eta_k - N\tilde{\Omega}/2+(N-1)\frac{V_1}{4}.
\end{align}
The system's ground state is still given
by $\ket{\tilde{G}} = \ket{--\dots -}$ and the orthogonal transformation of the operators preserves the
anti-commutator relations. Therefore, all other eigenstates can now
be constructed by the successive application of $\eta_k^\dagger$
operators for different $k$ since each mode can only be occupied by a
single excitation [$(\eta_k^\dagger)^2 =0$].
The matrix elements of $R$ and the transformed operators are given by
\begin{align}
  R_{jk}& = \sqrt{\frac{2}{N+1}}\sin\left(\frac{jk\pi}{N+1}\right), \\
 \eta_j &=  \sqrt{\frac{2}{N+1}}\sum_{k=1}^N \sin\left(\frac{jk\pi}{N+1}\right)c_k.
\end{align}
A straightforward calculation reveals that the eigenvalues of $M$
are given by $\lambda_j =
\tilde{\Omega}+\frac{V_1}{2}\cos(\frac{j\pi}{N+1})$.
Hence, our resulting Hamiltonian reads
\begin{align}
  \tilde{H}_{\rm xy} = \!\sum_{k= 1}^N  \eta_k^\dagger \eta_k
  \!\left[\tilde{\Omega}+\frac{V_1}{2}\cos\left(\!\frac{k\pi}{N+1}\right)\right]\!  -\! N\tilde{\Omega}/2+(N\!-\!1)\frac{V_1}{4}.
\end{align}
The fermionic excitation number operators $\eta_n^\dagger
\eta_n$ appearing in this Hamiltonian have only $0$ and $1$ as eigenvalues. Neglecting momentarily the term $-\! N\tilde{\Omega}/2+(N\!-\!1)\frac{V_1}{4}$,
the spectrum is thus given by 
\begin{align}
& \Bigg\{ E_{k_1k_2\dots k_l} =  \tilde{\Omega} \left[ l + \frac{V_1}{2\tilde{\Omega}}
  \sum_{n=1}^l\cos\left(\!\frac{k_n\pi}{N+1}\right) \right];  \nonumber \\ 
&\quad l \in
\{1,\dots, N\},k_1<k_2<\dots < k_l \in \{1,2,\dots, N\}\Bigg\}.
\end{align}
The corresponding basis states are given by $\ket{k_1k_2\dots k_l} := \eta^\dagger_{k_1}\eta^\dagger_{k_2}\cdots\eta^\dagger_{k_l}\ket{\tilde{G}}$.
Note that our total excitation number operator obeys
\begin{align}
\tilde{N}_+ & = \sum_{k=1}^N \sigma_+^{(k)}\sigma_-^{(k)} =
\sum_{k=1}^N c_k^\dagger c_k \\
& = \sum_{k,n,m = 1}^N
R_{nk}R_{mk}\eta_n^\dagger\eta_m =\sum_{n=1}^N \eta_n^\dagger\eta_n  
\end{align}
and counts the $\eta$-mode excitations:
\begin{align}
\tilde{N}_+\ket{k_1k_2\dots k_l} = l\ket{k_1k_2\dots k_l}.
\end{align}
This confirms that the neglected second part of the original perturbation 
$ \tilde{H'}_2 = \frac{V_1}{4} \sum_{k=1}^{N-1} [\sigma_+^{(k)}\sigma_+^{(k+1)}+
 \sigma_-^{(k)}\sigma_-^{(k+1)}]$ is fully off-diagonal in our
final eigenbasis, as it maps any basis state with $l$ excitations into
a superposition of states containing $l \pm 2$ excitations. 
Hence, for $\tilde{\Omega}\gg V_1$ the off-diagonal
perturbation only contributes to the spectrum at second order
$O(V_1^2/\tilde{\Omega})$ and to the eigenstates at order $O(V_1/\tilde{\Omega})$, justifying our former approach.

 \begin{figure}
  \includegraphics[width=8cm]{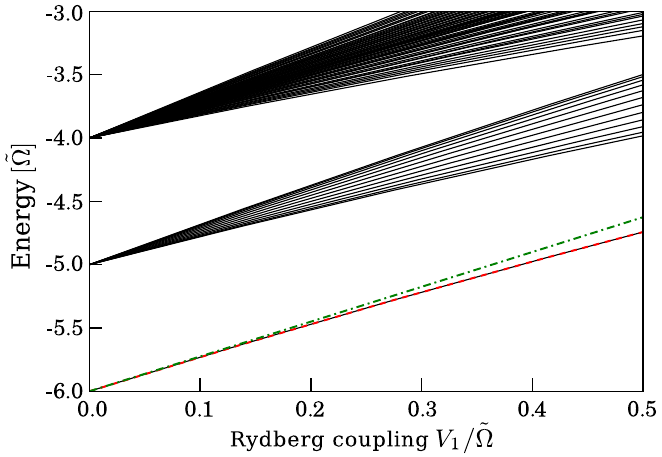}
      \caption{
Energy spectrum in the strong laser regime, showing the ground state, the $N_+= 1$
        subspace, and partially the $N_+= 2$ subspace for $N=12$. Without the effect of the perturbation $\tilde{H'}_2$, the energy
        levels would scale linearly with $V_1$ for fixed
        $\tilde{\Omega}$. Here, we see that for large perturbations,
        the linear scaling is slightly violated. For the ground level we have
        also plotted the analytical result of the unperturbed energy
        level (green, dash-dotted line) as well as the result of
        second order perturbation theory (red, dashed line).}
 \label{fig:laser_dominant_perturbing_interactions}
 \end{figure}

For the ground state and the first excited states we have explicitly
calculated the resulting second order energy corrections (see appendix).
Figure~\ref{fig:laser_dominant_perturbing_interactions} presents the
numerically calculated spectrum of $H$ as well as the analytically
calculated results for the ground state. A comparison of the
analytical and numerical results for the energy eigenvalues of the
$\tilde{N}_+ = 1$ subspace $\{\eta_k^\dagger\ket{G},\;k=1,2,\dots,N\}$ is
provided in table \ref{tab:strong_laser_energy_corrections}. 
In general, a very good agreement between the analytical and numerical results is obtained, demonstrating that $\tilde{H}_b$ and $\tilde{H}'_2$ only lead to minor corrections.

\begin{table}
\begin{ruledtabular}
\caption{\label{tab:strong_laser_energy_corrections}Comparison of the
 $N_+=1$ energy levels calculated analytically from the unperturbed $H_{XY}$
 Hamiltonian with results obtained via second order perturbation
 theory and the exact numerical results for $N=12$ and
 $V_{1}/\tilde{\Omega} = 0.1$.}
\begin{tabular}{c c c c}
$k$ & $E_k^{(0)}$ & $E_k^{(0)}+\Delta E_k^{(2)}$ & $E_k^{\rm num}$ \\
\hline
 1 & -4.676 & -4.681 & -4.681 \\
 2 & -4.681 & -4.685 & -4.685 \\
 3 & -4.688 & -4.691 & -4.692 \\
 4 & -4.697 & -4.701 & -4.700 \\
 5 & -4.707 & -4.709 & -4.711 \\
 6 & -4.719 & -4.723 & -4.722 \\
 7 & -4.731 & -4.733 & -4.734 \\
 8 & -4.743 & -4.746 & -4.746 \\
 9 & -4.753 & -4.756 & -4.757 \\
 10 & -4.762 & -4.767 & -4.766 \\
 11 & -4.769 & -4.774 & -4.774 \\
 12 & -4.774 & -4.778 & -4.778
\end{tabular}
\end{ruledtabular}
\end{table}

For completeness, we give a brief discussion of the symmetry
properties of the eigenstates. A detailed calculation reveals that the transformation properties under
reflection are
\begin{align}
  \mathcal{R}^{\dagger}\eta^{\dagger}_{k}\mathcal{R} =
  (-1)^{k+1}\eta^{\dagger}_{k}(-1)^{\tilde{N}_{+}} =   (-1)^{\tilde{N}_{+}+k}\eta^{\dagger}_{k}.
\end{align}
For a single fermionic excitation
$\eta^{\dagger}_{k}\ket{\tilde{G}}$, its eigenvalue
with respect to the reflection symmetry operator is correspondingly given by
$(-1)^{k+1}$.
Hence, only states corresponding to an odd
$k=1,3,5,\dots$ belong to the same symmetry eigenspace as the
canonical product ground state.
For a general state $\eta^\dagger_{k_1}\eta^\dagger_{k_2}\cdots \eta^\dagger_{k_l} \ket{\tilde{G}}$ the eigenvalue is given by $(-1)^{k_1+k_2+\dots k_l + l(l+1)/2}$. 

In concluding this section, we remark that the results
obtained here only apply in a rotating frame of reference due to the RWA
picture. The expectation value with respect to a given state $\ket{\psi}$ of
any operator $\hat{O}$ diagonal in the canonical product basis
$\mathcal{S}$ can be evaluated as usual:
$  \langle \hat{O} \rangle = \bra{\psi}\hat{O} \ket{\psi}$.
Any off-diagonal operator, on the other hand, must first be transformed to the
RWA frame and therefore becomes explicitly time dependent. Within the weak laser regime discussed in the previous section, most of the relevant observables are diagonal in the canonical product basis and are thus not affected by the RWA-picture.
For the strong laser regime eigenbasis, the situation is more complicated and needs to be inspected individually.

\subsection{Intermediate Regime}
\label{sec:intermediate_regime}
\begin{figure}
	\centering
	\includegraphics[width=8cm]{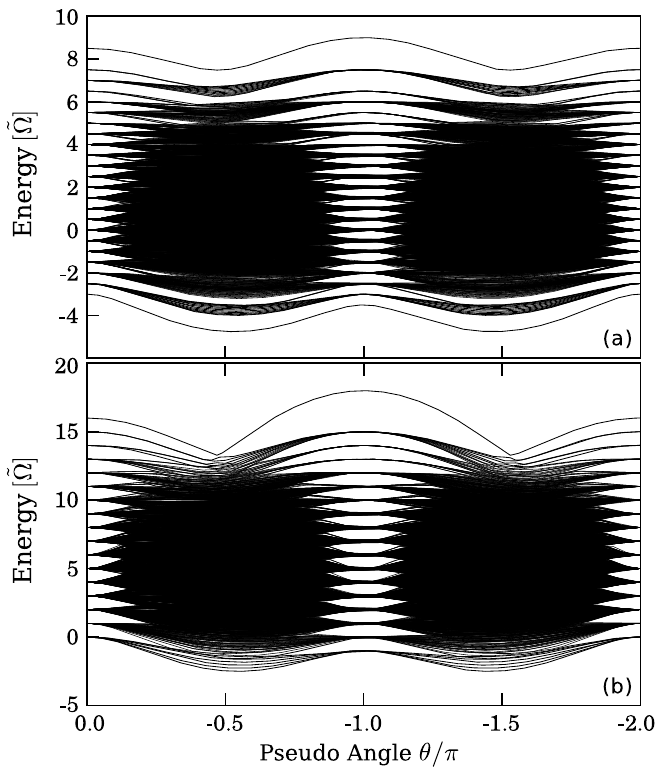}
	\caption{\label{fig:intermediate_regime} 
Energy spectrum obtained in the intermediate regime as a function of varying $\theta$ while
keeping the Rydberg interaction energy and the pseudo laser coupling constant. (a) $V_1/\tilde\Omega=1/2$ and (b) $V_1/\tilde\Omega=2$.
The clear separation of the energy bands as seen for the strong laser regime in figure \ref{fig:laser_dominant_varying_theta_2} is lost.
          }
\end{figure}
In the intermediate regime, where our parameters $\Delta, \Omega,
V_{1}$ can be of the same magnitude, we cannot separate 
the energy scales due to Rydberg interactions and of the laser contributions. 
We will therefore limit ourselves to the discussion of some
numerically obtained results. 
As in the previous section, we fix the relative scale of Rydberg
interaction energies and laser contributions by varying the
pseudo angle $\theta$ for fixed $\tilde\Omega$ and $V_1$.
Figure \ref{fig:intermediate_regime} provides two examples of the
resulting spectrum, namely for $V_1=\tilde\Omega/2$ and $V_1=2\tilde\Omega$, respectively. 
For $\theta = 0, \pi, 2\pi$, the Hamiltonian is
diagonal in the canonical product basis $\mathcal{S}_{N}$ and the
spectra can be fully understood from our discussion of the weak laser
regime.
For intermediate values of $\theta$, however, our analytical results
for the strong laser regime break down because the mixing of the
energy bands due to the neglected part of the interactions and the
boundary term eventually prevail. Accordingly, in figure \ref{fig:intermediate_regime} the corresponding energy bands cannot be resolved anymore and only the lowest and highest energy states remain well separated.

\section{Non-uniform Lattices}
\label{sec:patterned_lattices}
In this section we generalize our setup to allow for lattices with variable lattice spacings.
Due to the spatial dependence of the inter-site interactions this introduces new energy scales into the spectrum.
To account for the different lattice spacings, we define the spatially
dependent lattice spacing $a^{(k)}$ to be the distance between the lattice
sites $k$ and $k+1$.
Our interaction coefficients are then given by
\begin{align}
	\mathcal{V}_{k,k+l} &  =  \frac{C_n}{\left[\sum_{j=1}^{l} a^{(k + j -1)}\right]^n}.
\end{align}

\subsection{Alternating Spacings}
\label{sec:alternating_spacings}
We initially consider the specific case of a lattice with two alternating spacings $a_1,a_2$, such that the lattice sites are effectively grouped in pairs separated by the smaller spacing. For an even number of lattice sites this preserves the reflection symmetry of the Hamiltonian. 
For simplicity we introduce the ratio $\Gamma =
a_1/a_2$ between the lattice spacings.
Then, the next-neighbour interactions are given by
$V := C_6/a_1^6$ and $W := C_6/a_2^6 = \Gamma^{6} V$, respectively.
For $\Omega/V \ll 1$ we can again realize a weak laser coupling regime that allows for an analytical investigation similar to Sec.~\ref{sec:weak_laser_regime}. 
We split up the Hamiltonian $H = H_0 + H_L + H_{\rm int}^{\rm lr}$ into a dominant contribution $H_0 = H_{\rm int}^{\rm nn} + H_D$ given by the familiar  laser detuning operator and the next-neighbour Rydberg interactions and a perturbation given by the comparatively small laser coupling Hamiltonian and the long range Rydberg interactions $H_{\rm int}^{\rm lr}$.
For an even number of lattice sites, the next-neighbour interactions may be written as
\begin{align*}
        H_{\rm int}^{\rm nn} & = V \sum_{k=1}^{N/2} n_e^{(2k-1)} n_e^{(2k)} 
        + W  \sum_{k=1}^{N/2}n_e^{(2k)} n_e^{(2k+1)} \\
& =  V \left( N_{ee}^V  +\Gamma N_{ee}^W\right).
\end{align*}
Here, we have already divided the next-neighbour interactions $H_{\rm int}^{\rm nn}$
into pairs separated by $a_1$ and $a_2$, respectively, by introducing the corresponding next-neighbour pair operators $N_{ee}^V =\sum_{k=1}^{N/2}n_e^{(2k-1)} n_e^{(2k)}$ and $N_{ee}^W = \sum_{k=1}^{N/2}n_e^{(2k)} n_e^{(2k+1)}$, respectively.
In the above-mentioned case of a lattice with an odd number of sites, these two operators must be appropriately modified, but most of the general spectral features are similar.
The unperturbed Hamiltonian $H_0$ is thus given by
\begin{align}
	H_0 & = \Delta \left( N_e - N/2\right) + V\left[ N_{ee}^{V} +  \Gamma^6N_{ee}^{W} \right].
\end{align}
Since $H_0$ is diagonal in the canonical product basis, this again leads to a simple formula for the eigenvalue of $H_0$ for a canonical basis state $\ket{S} \in \mathcal{S}_\mathcal{H}$,
\begin{align}\label{eq:energy_ne_nee_nee}
	E^{\Gamma}(S) & = \Delta\left[N_e(S) - N/2\right] + V \left[N_{ee}^{V}(S) + \Gamma^6 N_{ee}^{W}(S) \right],
\end{align}
where $N_{ee}^{V}(S)$ and $N_{ee}^{W}(S)$ are the
eigenvalues of the excitation pair operators $N_{ee}^{V}$ and
$N_{ee}^{W}$ for the state $\ket{S}$.

\begin{figure}
  \centering    
  \includegraphics[width=8cm]{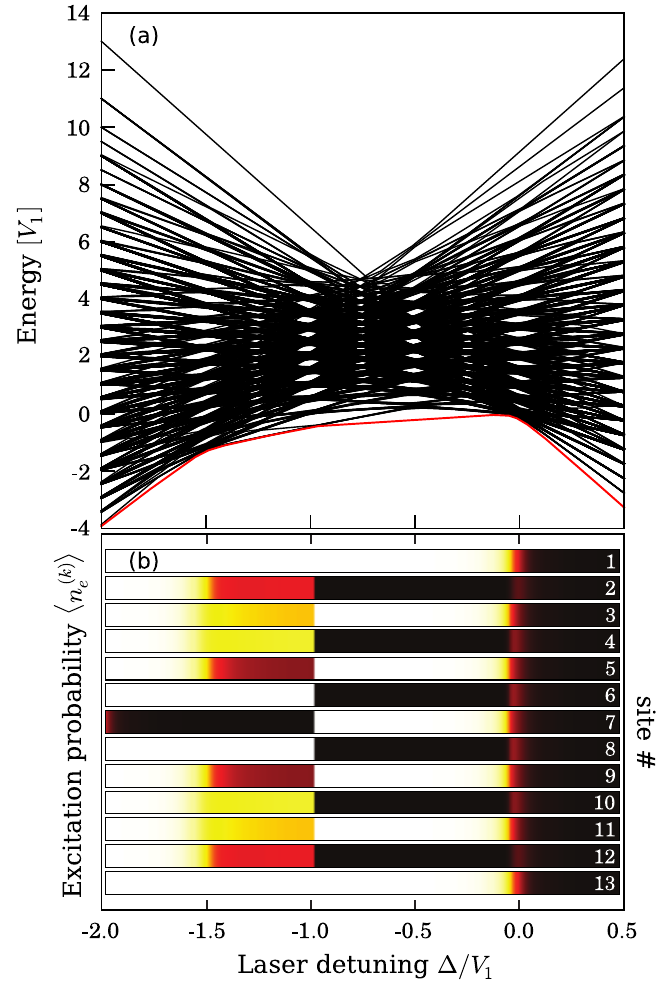}
\caption{\label{fig:spectra_patterned_weak_omega_varying_delta}
 (a) Spectrum and (b) local excitation probability for the energetic ground state in a patterned lattice with $N= 13$, $\Gamma = \frac{a_1}{a_2} = \sqrt[6]{2}$ and $\Omega = 0.05V$. In contrast to the case of a single lattice spacing (cf.~figure \ref{fig:spectra_weak_omega_varying_delta}) we see that for $\Delta \in [-2 V, -V]$ the ground state is given by superpositions of states featuring different excitation patterns, especially for $\Delta/V<-1$. In panel (b) white regions correspond to unity Rydberg excitation probability.
 }
\end{figure}
Figure \ref{fig:spectra_patterned_weak_omega_varying_delta} presents
the resulting energy spectrum and excitation pattern of the ground state
obtained for $\Gamma^{6} = 2$ and an odd number of lattice sites $N=13$. Comparing this
result with the case of a constant spacing (equivalent to $\Gamma=1$)
shows a qualitatively different excitation pattern for the energetic ground state for $\Delta\le-V$. 
Formally, the resulting energy eigenvalues [cf.~\eqref{eq:energy_ne_nee_nee}] share some similarities with our result obtained
for a constant lattice spacing
[cf.~\eqref{eq:energy_ne_nee}]. $V$ plays the role of $V_1$ while
$N_{ee}^{\rm eff} := \left[N_{ee}^{V}(S) + \Gamma^6 N_{ee}^{W}(S) \right]$ may
be understood as an effective, in our case of $\Gamma^6 =2$ integer, pair number. It is clear that, depending
on the specific value of $\Gamma^6$, this generalization allows for
more kinds of degeneracies as observed in figure \ref{fig:spectra_patterned_weak_omega_varying_delta}(a). Although there exist further combinatorial
constraints on $N_{ee}^{W}(S)$ and $N_{ee}^{V}(S)$
(e.g., $N_{ee}^W + N_{ee}^V = N_{ee}$) it is now possible to bring two
states featuring the same number of $N_e$ excitations but different
excitation pair numbers into degeneracy.
This cannot be achieved dynamically, as this would require a
variation of the lattice constants during the experiment, which is likely to destroy the quantum
coherence of the system. But it may be exploited to adiabatically
prepare superpositions of lattice states with different pair
numbers $N_{ee}^W, N_{ee}^V$ by means of a time-dependent detuning
$\Delta(t)$.

We also remark at this point that the results obtained so far already have an interesting consequence: distinct spectral features will depend in some fashion on the sixth power of the ratio of the lattice spacings $\Gamma^6$. This exponent is due to the assumed van-der-Waals interaction potential which scales with the interatomic distance $R$ as $R^{-6}$. For a general scaling $\mathcal{V}(R) \propto R^{-n}$ the ratio would enter as $\Gamma^n$. Hence, an experimental study of such a system could exploit this to verify the spatial dependence of the interaction potential. This would be achieved by choosing values for $\Delta$, $V$ and $\Gamma$ which lead to very specific and ideally experimentally testable degeneracies. However, this goes beyond the scope of the present work and will be subject of a separate study.

\subsection{Crystal State Transitions}
\label{sec:perfect_crystal_transitions}

\begin{figure}
  \centering
    \includegraphics[width=8cm]{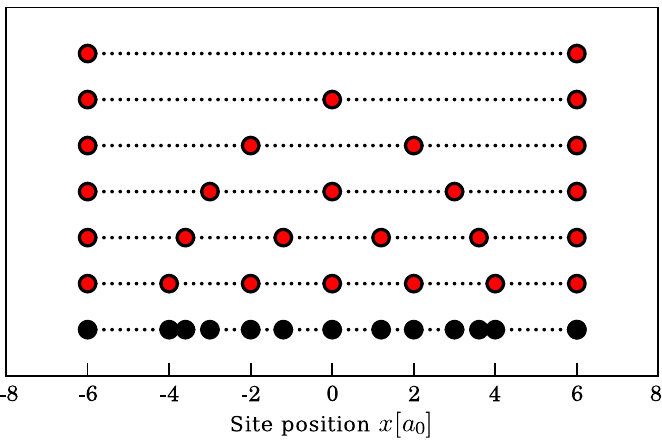}
\caption{\label{fig:n_61_pattern} The crystal states $\ket{[N_{e}]}$
  (excitations marked in red)
  are presented for $N_{e}=2,3,4,5,6,7$ in a lattice of size
  $N_{0}=61$. As can be seen, to realize these states, only $N=13$ lattice
sites (lowest row) must actually be occupied by atoms.}
\end{figure}
\begin{figure}
  \centering
    \includegraphics[width=8cm]{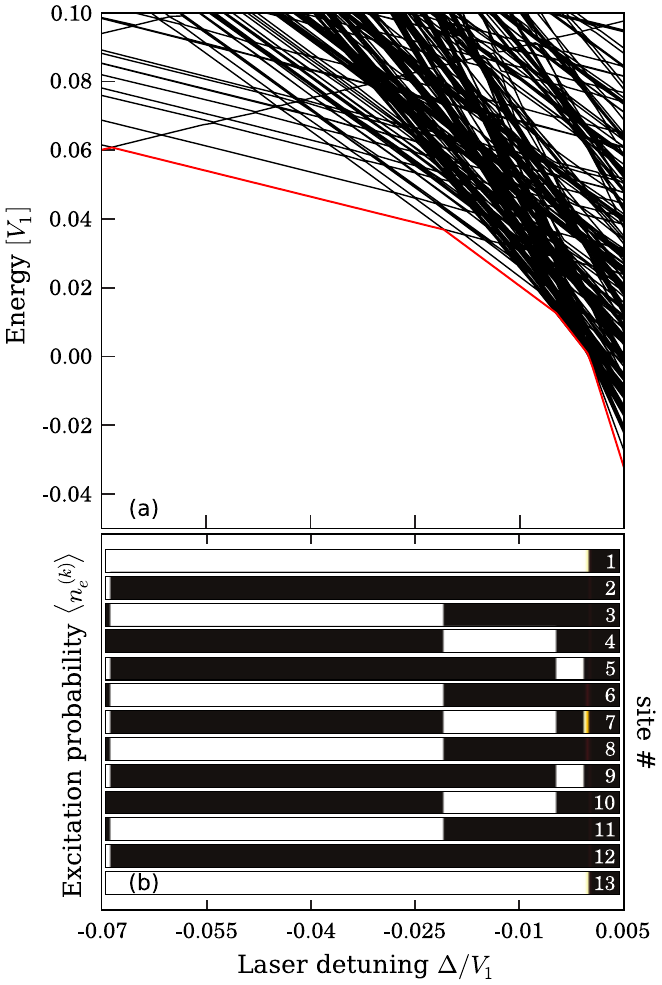}
\caption{\label{fig:spectra_patterned_perfect_transitions}
(a) Spectrum and (b) local excitation probability for the energetic ground state in the patterned lattice of figure \ref{fig:n_61_pattern} with $\Omega = 10^{-4}V$. As discussed in the text, this particular setup emulates a $N_0=61$ site lattice. The transitions between the crystal states $\ket{[N_e]}$ are evident in panel (b) where white regions correspond to unity Rydberg excitation probability.}
\end{figure}
A further interesting sequence of ground state transitions follows from our previous
investigation of the succession of crystal states in the full blockade
regime.
For $N=13$ we can realize pronounced transitions
$\ket{[2]}\to\ket{[3]}\to\ket{[4]}\to\ket{[5]}$ for the case of a single-spaced lattice (see 
section \ref{sec:single_spaced_lattices}). For a lattice of
size $N_{0}= 61$, on the other hand, we would be able to achieve two additional transitions
$\ket{[2]}\to\ket{[3]}\to\ket{[4]}\to\ket{[5]}\to\ket{[6]}\to\ket{[7]}$. 
However, the energy spacing between the crystal states scales with $(N-1)^{-6}$ for a fixed number of excitations, rendering an experimental realization impractical. This drawback can be compensated, though, by decreasing the lattice constant $a_0$.

There exists an alternative possibility of realizing the $N = 61$
crystal state transitions with a strongly non-uniform lattice consisting of only $N=13$
lattice sites. In principle this can be thought of a full
$N=61$ lattice, where the atoms have been removed from $48$ sites. Figure \ref{fig:n_61_pattern}
illustrates the corresponding spatial pattern according to which the sites have to be
arranged. For the case of figure \ref{fig:n_61_pattern}, the values of the detunings at the transition points
are given by \eqref{eq:transition_delta_exact} with
$N=13$, providing a more favourable scaling of the energy spacing between the crystal states.
As indicated above, the same transition detunings can be achieved by a uniform $N=61$ lattice with a reduced lattice spacing of $0.2\, a_0$. Here, $a_0$ denotes the spacing of a uniform lattice consisting of $N=13$ sites.
The redistributed $N=13$ lattice, on the other hand, only demands a minimal spacing of $0.4\, a_0$. If the minimal lattice spacing is experimentally limited, the redistributed $N=13$ lattice thus provides a more favourable scaling of the relevant detunings.

Figure \ref{fig:spectra_patterned_perfect_transitions}(a)
presents the energy spectrum with the first five transitions 
between crystal states. In panel (b) the corresponding local
Rydberg excitation probability is illustrated. One can nicely observe
the transitions between the crystal states.
In order to get the sharp transitions between the crystal states, a relatively small Rabi frequency of $\Omega=10^{-4}V$ is assumed in this figure.
Due to the sensitive scaling of the corresponding transition detuning values with the interaction
exponent $n$, their exact experimental measurement could provide a
useful test of the validity of the assumed interaction potential.

\section{Brief Summary} 
\label{sec:conclusions_and_outlook}\label{sec:app_strong_laser}

Focusing on the frozen gas regime at ultracold temperatures, we have provided a comprehensive analysis of the spectral properties of the coherent, laser-driven Rydberg excitation in one-dimensional ordered systems.
In the weak laser regime, a systematic classification of the state space has been performed. In the so-called full blockade regime, where no neighbouring Rydberg excitations in the lattice are allowed, the transitions between crystal Rydberg lattice states were identified. 
Turning our investigations to the strong laser regime, an extensive analytic treatment allowed the full characterization of the spectrum by performing site-specific rotations in spin space and introducing fermionic ladder operators. An excellent agreement of our perturbative treatment and the numerically obtained spectrum has been found.
Going from regularly spaced to patterned lattices, we investigated two particular examples. For alternating lattice constants, an additional energy scale is introduced that alters the spectral properties as well as the local Rydberg excitation probability of the ground state. In addition, we have highlighted a specific setup involving an irregular lattice site distribution that allows the emulation of the crystal state transitions of a more extended one-dimensional lattice.

\begin{acknowledgments}
We are very grateful to Wolfgang Zeller for both fruitful discussions and practical help concerning technical aspects of this work. 
We thank I. Lesanovsky, B. Olmos and T. Pohl for insightful discussions. M.M. acknowledges financial support from the German Academic Exchange Service (DAAD).
\end{acknowledgments}

\begin{appendix}
\section{Perturbative Energy Corrections in the Strong Laser Regime}
Here we present the formulas used for calculating the perturbative energy corrections presented in Section \ref{sec:strong_laser_regime}.
This is achieved by re-expressing the perturbation operators in terms of the fermionic ladder operators $\eta_k  = \sum_{l=1}^{N} R_{kl} c_l$. 
The coefficient matrix $(R_{kl})_{k,l=1}^N$ is real, symmetric and orthogonal:
\begin{align*}
    & R_{kl} = R_{lk} = \sqrt{\tfrac{2}{N+1}} \sin \tfrac{kl\pi}{N+1}, \\
    & \sum_{l=1}^N R_{kl}R_{jl} = \sum_{l=1}^N R_{kl}R_{lj} = \delta_{kj}.
\end{align*}
This can be proved by means of the following identities.
We find that for $m \in \mathbb{Z}$
\begin{align*}
    \sum_{n=0}^N \exp(i\tfrac{nm\pi}{N+1}) = 
        \begin{cases} N+1 &\text{ for } m \in 2(N+1)\mathbb{Z}, \\
                    1 + i \frac{1 + \cos (\tfrac{m\pi}{N+1}) }{\sin (\tfrac{m\pi}{N+1})} &\text{ for odd } m, \\
                    0 & \text{ otherwise}.
        \end{cases}
\end{align*}
From this, it is straightforward to show that for $k,l \in \{1,2,\dots, N\}$
\begin{align*}
    &\sum_{n=1}^{N}\sin (\tfrac{nk\pi}{N+1}) \sin(\tfrac{nl\pi}{N+1}) \\
    = &-\frac{1}{2} \Re \sum_{n=0}^N \left\{ \exp [i\tfrac{n(k+l)\pi}{N+1}] - \exp [i\tfrac{n(k-l)\pi}{N+1}] \right\} \\
    = &\frac{N+1}{2}\delta_{kl},
\end{align*}
and one finds therefore
\begin{align*}
    & \sum_{n=1}^{N}\sin (\tfrac{nk\pi}{N+1}) \cos(\tfrac{nl\pi}{N+1}) \\
    = & \frac{1}{2} \Im \sum_{n=0}^N \left\{ \exp [i\tfrac{n(k+l)\pi}{N+1}]  + \exp [i\tfrac{n(k-l)\pi}{N+1}] \right\} \\
    = & \begin{cases} 0 & \text{ for even } (k \pm l) \\
         \frac{\sin(\tfrac{k\pi}{N+1})}{\cos(\tfrac{l\pi}{N+1}) - \cos(\tfrac{k\pi}{N+1})} & \text{ for odd } (k\pm l)
    \end{cases}
\end{align*}
Where $\Re, \Im$ indicate the real and imaginary part. The unperturbed XY-Hamiltonian reads
\begin{align}
  \tilde{H}_{\rm xy} &= \sum_{k= 1}^N  \eta_k^\dagger \eta_k
  \left[\tilde{\Omega}+\frac{V_1}{2}\cos(\frac{k\pi}{N+1})\right] + \text{const}.
\end{align}
and the perturbations are given by the boundary term
\begin{align}
 \tilde{H}_b = \frac{V_1}{4}\left[\sigma_x^{(1)} + \sigma_x^{(N)}\right]
\end{align}
and the off-resonant couplings
\begin{align}
    \tilde{H'}_2 = \frac{V_1}{4} \sum_{k=1}^{N-1} \left(\sigma_+^{(k)}\sigma_+^{(k+1)}+
     \sigma_-^{(k)}\sigma_-^{(k+1)}\right).
\end{align}
The first energy manifold is spanned by the ground state $\ket{\tilde{G}}$. 
The first excited energy manifold is given by the set of states that contain a single excitation
\begin{align*}
    \left\{ \eta_k^\dagger \ket{\tilde{G}},\; 1 \le k \le N\right\}.
\end{align*}
One finds that
\begin{align*}
    \tilde{H}_b & = \frac{V_1}{4}\left[ c_1^\dagger + c_1 + (c_N^\dagger - c_N)(-1)^{\tilde{N}_+} \right]\\
    & = \frac{V_1}{4}\sum_{k=1}^N R_{1k}\left\{ \left[1 + (-1)^{k + \tilde{N}_+}\right]\eta_k^\dagger  + \left[1 - (-1)^{k + \tilde{N}_+}\right]\eta_k \right\},
\end{align*}
where we have used $R_{Nk} = (-1)^{k+1} R_{1k}$.
The second part of the perturbation is
\begin{align}
    \tilde{H'}_2 & = \frac{V_1}{4} \sum_{n=1}^{N-1} \left(c_n^\dagger c_{n+1}^\dagger - c_n c_{n+1}\right) \\  
    & = \frac{V_1}{4} \sum_{k,l=1}^{N} \left(\sum_{n=1}^{N-1} R_{k,n}R_{l,n+1} \right) \left(\eta_k^\dagger \eta_l^\dagger - \eta_k \eta_l\right).
\label{eq:pert_off_eta}
\end{align}
Since the last term in \eqref{eq:pert_off_eta} is antisymmetric with respect to $k$ and $l$ and they are summed over their full range, it is convenient to calculate the antisymmetrized value of the bracketed expression. We define thus
\begin{align*}
    F_{kl} & := \half \left[ \left( \sum_{n=1}^{N-1} R_{k,n}R_{l,n+1} \right) - (k \leftrightarrow l) \right] \\
        & = \tfrac{1}{N+1}\sum_{n=0}^{N} \left\{\sin (\tfrac{n k \pi}{N+1}) \sin [\tfrac{(n+1) l \pi}{N+1}] - (k \leftrightarrow l)\right\} \\
        & = \tfrac{1}{N+1}\sum_{n=0}^{N} \left\{ \left[ \sin (\tfrac{n k \pi}{N+1}) \sin (\tfrac{n l \pi}{N+1}) \cos (\tfrac{ l \pi}{N+1}) \right.\right. \\
        &  \quad + \left.\left.\sin (\tfrac{n k \pi}{N+1}) \cos (\tfrac{nl \pi}{N+1})\sin (\tfrac{l \pi}{N+1})\right] - (k \leftrightarrow l)\right\} \\
        & = \begin{cases} 0 & \text{ for even } (k \pm l), \\ 
            \frac{2}{N+1} \frac{\sin (\tfrac{k\pi}{N+1}) \sin (\tfrac{l\pi}{N+1})}{\cos (\tfrac{l\pi}{N+1})- \cos (\tfrac{k\pi}{N+1})} & \text{ for odd } (k \pm l).
            \end{cases}        
\end{align*}
The off-resonant couplings are then given by
\begin{align*}
    \tilde{H'}_2  & = \frac{V_1}{4} \sum_{k,l=1}^{N} F_{kl} \left(\eta_k^\dagger \eta_l^\dagger - \eta_k \eta_l\right).
\end{align*}
We thus see explicitly that the perturbation operators only couple manifolds of different numbers of fermionic excitations, specifically $\Delta \tilde{N}_+ = \pm 1$  for $\tilde{H}_b$ and $\Delta \tilde{N}_+ = \pm 2$ for $\tilde{H'}_2$, respectively. This implies that all first order energy corrections vanish due to the fact that $\tilde{N}_+$ is diagonal in our basis.
In our specific case we find that we can calculate the second order contributions due to $\tilde{H}_b$ and $\tilde{H'}_2$ independently, since for any given eigenstate these operators couple to different subspaces.
The relevant matrix elements for the corrections to the ground state are given by
\begin{align*}
    \bra{\tilde{G}}\eta_k \tilde{H}_b \ket{\tilde{G}} & = \frac{V_1}{4} R_{1k} \left[ 1 - (-1)^{k}\right], \\
    \bra{\tilde{G}}\eta_k\eta_l \tilde{H'}_2 \ket{\tilde{G}} & = - \frac{V_1}{2} F_{kl}.
\end{align*}
From these results and the specific form of $F_{kl}$ we see that only specific states contribute to the energy corrections. This is due to the overall symmetry under reversal of the lattice, which is preserved by the perturbations.
For the corrections to the states of the first excited manifold we need the following matrix elements
\begin{align*}
    & \bra{\tilde{G}}\eta_k\eta_l \tilde{H}_b \eta_q^\dagger \ket{\tilde{G}} \\ 
    & \qquad = \frac{V_1}{4} \left\{ R_{1l} \left[ 1 + (-1)^{l}\right] \delta_{kq} - R_{1k} \left[ 1 + (-1)^{k}\right] \delta_{lq}\right\}, \\
    & \bra{\tilde{G}}\eta_k\eta_l\eta_m \tilde{H'}_2 \eta_q^\dagger\ket{\tilde{G}}\\
    & \qquad = - \frac{V_1}{2} \left( F_{kl}\delta_{mq} - F_{km}\delta_{lq} + F_{lm}\delta_{kq}\right).
\end{align*}
The ground state energy correction is thus given by
\begin{align*}
    \Delta E_{\tilde{G}}^{(2)} & = \Delta E_{\tilde{G},b}^{(2)} + \Delta E_{\tilde{G},\rm off}^{(2)}, \\
    \Delta E_{\tilde{G}, b}^{(2)} & = - \frac{V_1^2}{4 \tilde{\Omega}} \sum_{k=1}^{\lceil N/2 \rceil} \frac{ R_{1,2k-1}^2 }{ 1 + \tfrac{V_1}{2\tilde{\Omega}} \cos\tfrac{(2k-1)\pi}{N+1}}, \\
    \Delta E_{\tilde{G},\rm off}^{(2)} & = - \frac{V_1^2}{8 \tilde{\Omega}} \sum_{k=1}^N\sum_{l=k+1}^N \frac{F_{kl}^2}{1 + \tfrac{V_1}{4\tilde{\Omega}} (\cos\tfrac{k\pi}{N+1} + \cos\tfrac{l\pi}{N+1})}.
\end{align*}
Before we continue by writing down the energy corrections to the singly excited states we should note that we can further approximate the energy corrections by neglecting the next to leading order contributions in the denominators:
\begin{align*}
    \Delta E_{\tilde{G}, b}^{(2)} & \approx - \frac{V_1^2}{4 \tilde{\Omega}} \sum_{k=1}^{\lceil N/2 \rceil} R_{1,2k-1}^2, \\
    \Delta E_{\tilde{G},\rm off}^{(2)} & \approx - \frac{V_1^2}{8 \tilde{\Omega}} \sum_{k=1}^N\sum_{l=k+1}^N F_{kl}^2.
\end{align*}
Making the same additional approximation as above, we find for the corrections to an excited state $\eta^\dagger_q \ket{\tilde{G}}$
\begin{align*}
    \Delta E_q^{(2)} & = \Delta E_{q,b,\tilde{G}}^{(2)} + \Delta E_{q,b}^{(2)} + \Delta E_{q,\rm off}^{(2)}, \\
    \Delta E_{q,b,\tilde{G}}^{(2)}& \approx \frac{V_1^2}{4 \tilde{\Omega}} R_{1,q}^2 \frac{1-(-1)^q}{2} , \\
     \Delta E_{q,b}^{(2)} & \approx - \frac{V_1^2}{4 \tilde{\Omega}} \mathop{\sum_{k=1}}_{k\neq q}^N R_{1,k}^2 \frac{1+(-1)^k}{2}\\
  \Delta E_{q,\rm off}^{(2)} & \approx - \frac{V_1^2}{8 \tilde{\Omega}}  \sum_{k=1}^N\left( \sum_{l= k+1}^N F_{kl}^2 -  F_{kq}^2\right).
\end{align*}
\end{appendix}

\bibliography{1dRydLattice}
\end{document}